\documentclass[fleqn,usenatbib]{mnras}

\usepackage{newtxtext,newtxmath}

\usepackage[T1]{fontenc}
\usepackage{anyfontsize}

\DeclareRobustCommand{\VAN}[3]{#2}
\let\VANthebibliography\thebibliography
\def\thebibliography{\DeclareRobustCommand{\VAN}[3]{##3}\VANthebibliography}

\usepackage{graphicx}	
\usepackage{amsmath}	

\usepackage{amssymb}	
\usepackage{multicol}
\usepackage{mathtools, cuted}
\usepackage{caption}
\usepackage{subcaption}
\usepackage{float}
\usepackage{verbatim}
\usepackage[section]{placeins}

\newcommand{\hompc}{\,h\,\mathrm{Mpc}^{-1}}

\DeclareMathOperator{\Tr}{Tr}
\DeclareMathOperator{\cov}{\boldsymbol{\mathsf{C}}}
\DeclareMathOperator{\icov}{\boldsymbol{\Psi}}
\DeclareMathOperator{\comp}{\boldsymbol{\mathsf{D}}}
\DeclareMathOperator{\win}{\boldsymbol{\mathsf{W}}}

\title[Fast cosmological analysis]{Faster cosmological analysis with power spectrum without simulations}

\author[Yan Lai et al.]{
Yan Lai,$^{1}$\thanks{E-mail: y.lai1@uqconnect.edu.au}
Cullan Howlett,$^{1}$
Tamara~M.~Davis$^{1}$
\\
$^{1}$School of Mathematics and Physics, The University of Queensland, QLD 4072, Australia\\
}

\date{Accepted XXX. Received YYY; in original form ZZZ}

\pubyear{2023}

\begin{document}
\label{firstpage}
\pagerange{\pageref{firstpage}--\pageref{lastpage}}
\maketitle

\begin{abstract}
Future surveys could obtain tighter constraints on the cosmological parameters with the galaxy power spectrum than with the Cosmic Microwave Background. However, the inclusion of multiple overlapping tracers, redshift bins, and more non-linear scales means that generating the necessary ensemble of simulations for model-fitting presents a computational burden. In this work, we combine full-shape fitting of galaxy power spectra, analytical covariance matrix estimates, the MOPED (Massively Optimised Parameter Estimation and Data compression) method, and the Taylor expansion interpolation of the power spectrum for the first time to constrain the cosmological parameters directly from a state-of-the-art set of galaxy clustering measurements. We find it takes less than a day to compute the analytical covariance while it takes several months to calculate the simulated ones. Combining MOPED with the Taylor expansion interpolation of the power spectrum, we can constrain the cosmological parameters in just a few hours instead of a few days. We also find that even without \textit{a priori} knowledge of the best-fit cosmological or galaxy bias parameters, the analytical covariance matrix with the MOPED compression still gives consistent cosmological constraints to within $0.1\sigma$ after two iterations. Therefore, the pipeline we have developed here can significantly speed up the analysis for future surveys such as DESI and Euclid.  
\end{abstract}




\section{Introduction}

In the past few decades, galaxy redshift surveys have become one of the most powerful ways to constrain cosmological parameters. Although their constraining power is still weaker than measurements from the Cosmic Microwave Background (CMB), future galaxy surveys have the potential to surpass the CMB constraints \citep{Carrasco_2012}. In a typical analysis pipeline, the redshift measurements from the galaxy surveys are converted to the two-point correlation function or its Fourier counterpart, the power spectrum. Recent developments in the theory of two-point clustering \citep{McEwen_2016, Simonovi__2018, Ivanov_2020, Tr_ster_2020, Valogiannis_2020, D_Amico_2021, Chen_2021, Noriega_2022} have allowed us to use galaxy redshift surveys to obtain precise and accurate constraints directly on the parameters of our cosmological model \citep{d_Amico_2020, Glanville_2022, Philcox_2022, Semenaite_2022}, akin to those presented using CMB data.

Within this procedure, under the assumption that the distribution of the two-point clustering is Gaussian, the likelihood function for comparing the data and model can be fully described by using simulations to construct a `brute-force' estimation of the covariance matrix of the statistic. Over the last decade, this process has been developed from multiple angles into an almost industrial process, including new methods/algorithms for producing the required numbers of accurate simulations (e.g., \citealt{White_2013, Chuang_2014, Howlett_2015, Angulo_2016}), methods to quickly estimate the two-point clustering (e.g., \citealt{He_2021, Keih_nen_2022}), and advances in how the modeling and fitting are done (e.g., \citealt{Audren_2012, Zuntz_2015, Brieden_2021}).

One caveat of using simulations as described above is that they also introduce noise into the covariance matrix estimation. It is important to note that when using simulations, what we actually have is an \textit{estimate} of the true covariance matrix drawn from a Wishart distribution. As shown in \cite{Hartlap_2006, Dodelson_2013} and \cite{Percival_2014}, to suppress the noise in our estimate and generate an unbiased inverse of the covariance matrix, a large number of simulations is required, proportional to the size of the data vector. Furthermore, the appropriate Bayesian approach is to marginalize over the uncertainty of the true covariance matrix, which modifies the form of the Gaussian likelihood \citep{Sellentin_2015, Percival_2021}. This is a bottleneck for future surveys that will attempt to measure the clustering down to smaller scales and with more redshift bins, so we would require more simulations. To overcome some of these limitations and generate the covariance matrix faster, semi-analytical \citep{Hamilton_2006, O_Connell_2016, Pearson_2016}, approximations \citep{monaco2016approximate, Blot_2019}, and analytical \citep{Sugiyama_2020, Wadekar_2020} covariance matrices have been developed. These methods dramatically reduce the number of simulations required to generate an unbiased and accurate covariance matrix or provide alternative methods to compute the true covariance matrix without simulations. 

Alternatively, we can use data compression to reduce the size of the data vector such that we need fewer simulations to estimate its covariance matrix \citep{Heavens_2000, Cannon_2010, Gualdi_2019}. These compression methods can also potentially speed up the data analysis because the compressed covariance matrix and data vector are much smaller. Furthermore, with recent advancements in machine learning, several emulators such as \textsc{Bacco} \citep{Pellejero_Iba_ez_2023} and \textsc{matryoshka} \citep{Donald_McCann_2022} significantly speed up the power spectrum computation. The data analysis bottleneck becomes the chi-squared computation when using the emulators. In this case, the data compression can help to speed up the analysis because the size of the compressed data vectors and covariance matrix are smaller. 

In this work, we demonstrate that all of these concepts can be brought together to produce an accurate and fast estimate of cosmological parameters from galaxy redshift surveys with minimum numbers of simulations. We will focus on the analytical covariance matrix from \citet{Wadekar_2020}, the Massively Optimized Parameter Estimation and Data compression (MOPED) algorithm from \citet{Heavens_2000}, and the Taylor expansion interpolation of the power spectrum in \citet{Colas_2020}. All three methods have been applied to power spectrum analysis before \citep{Wadekar_2021, Gualdi_2019B, Colas_2020}. However, this is the first time that the combination of all three methods is used to analyze the power spectrum alongside modeling pipelines that allow us to fit directly for cosmological parameters. In this work, we will use the Python code for Biased tracers in redshift space (\textsc{PyBird}; \citealt{D_Amico_2021}) to calculate the power spectrum using the Effective Theory of Large-Scale Structure; EFTofLSS. A common argument against using Full-Shape modelling such as \textsc{PyBird} directly on the two-point clustering is that the large numbers of data points and free parameters used, particularly when fitting multiple redshift bins simultaneously, require a prohibitively large number of simulations for covariance matrix estimation. We can show that this is not the case by combining the above techniques in this work. 


This paper is organized as follows. In Section \ref{sec:motivation}, we provide further motivation for this work with simple numerical exercises and a case study involving the Dark Energy Spectroscopic Instrument (DESI; \citealt{DESI_2019}). Turning then to real data, Section~\ref{sec:data} introduces the combination of redshift data from the six-degree Field Galaxy Survey (6dFGS; \citealt{Jones_2009}), the Baryon Oscillation Spectroscopic Survey (BOSS; \citealt{Alam_2017}) and the extended Baryon Oscillation Spectroscopic Survey (eBOSS; \citealt{Alam_2021}) that we will utilize in this work. In Section~\ref{sec:theory}, we introduce the theory behind the analytical covariance matrix, the MOPED compression algorithms, and the Taylor expansion interpolation of the power spectrum. We present our results in Section~\ref{sec:results} before concluding in Section~\ref{sec:conclusion}.

\section{Motivation}
\label{sec:motivation}
Longer data vectors, or covariance matrices estimated from a small number of simulations, lead to increasingly biased and inaccurate likelihood estimation. In the simple case of a multi-variate Gaussian likelihood, \citet{Hartlap_2006} and \citet{Taylor_2013} developed various analytical formulae to measure the accuracy of the covariance $\cov$ and precision matrices $\icov = \cov^{-1}$. For example, the fractional bias can be written \citep{Taylor_2013} as
\begin{equation}
    B = \frac{\Tr \langle \icov \rangle}{ \Tr \hat{\icov}} = \frac{N_S-1}{N_S - N_D - 2} - 1 = H_p^{-1}-1,
    \label{eq:bias_C}
\end{equation}
where the unbiased (and unknown) true precision matrix is given by \(\hat{\icov}\) and the precision matrix as measured from simulations is given by \(\icov\). \(N_S\) denotes the number of simulations, and \(N_D\) is the length of the data vector.\footnote{There is a typo in equation (35) of \citet{Taylor_2013}, the right-hand side of the equation is missing "-1".} Therefore, to recover an unbiased estimate of the precision matrix, we need to multiply it by the Hartlap factor, $H_p$. 

The fractional error of the precision matrix is given by \citep{Taylor_2013}
\begin{equation}
    E_{\icov} = \sqrt{\frac{\Tr \sigma^2(\icov)}{\Tr (\hat{\icov})^2}} = H^{-1}_{p} \sqrt{\frac{2}{N_S-N_D-4}},
    \label{eq: frac_Cinv}
\end{equation}
where \(\sigma(\icov)\) denotes the standard deviation of the simulated precision matrices.\footnote{Our equation here is different from equation (36) in \citet{Taylor_2013} because our expression is valid for the precision matrix without being corrected by the Hartlap factor. Therefore, we get an extra Hartlap factor in the denominator. Equation (36) in \citet{Taylor_2013} also has a typo, the denominator should be \(N_S-N_D-4\) instead of \(N_S-N_D-2\). We derived equation~(\ref{eq: frac_Cinv}) using their equations (26) and (27).} The fractional error of the covariance matrix is given similarly \citep{Taylor_2013},
\begin{equation}
    E_{\cov} = \frac{\Tr \sigma(\cov)}{\Tr \hat{\cov}} = \sqrt{\frac{2}{N_S-1}}.
    \label{eq: frac_C}
\end{equation}

We can also measure the sampling noise of each element of the covariance matrix with \citep{Taylor_2013, Wadekar_2021}
\begin{align}
    \label{eq:delta_C}
    \Delta C_{l_1, l_2}(k_i, k_j) &= \frac{1}{\sqrt{N_s - 1}} \nonumber \\
    &\left(C^2_{l_1, l_2}(k_i, k_j)+ C_{l_1, l_1}(k_i, k_i)C_{l_2, l_2}(k_j, k_j)\right)^{\frac{1}{2}}
\end{align}
assuming the variations of the power spectra from the mock catalog are Gaussian distributed. Here, \(l_1\) and  \(l_2\) denote the multipole of the covariance matrix. 

Lastly, \cite{Percival_2014} demonstrated that the noise in the covariance matrix should also enlarge parameter constraints obtained using that covariance matrix. They provided an expression for increasing the quoted uncertainty to take this into account if a full Bayesian marginalization over this uncertainty has not been performed (i.e., \citealt{Sellentin_2015}). The appropriate factor in this case is
\begin{equation}
m_{1} = \frac{1 + B(N_{D} - N_{P})}{1+A+B(N_{P}+1)}
\label{eq:m1}
\end{equation}
where 
\begin{align}
A &= \frac{2}{(N_{S}-N_{D}-1)(N_{S}-N_{D}-4)} \notag \\
B &= \frac{N_{S}-N_{D}-2}{(N_{S}-N_{D}-1)(N_{S}-N_{D}-4)}
\end{align}
and $N_{P}$ is the number of free parameters in our model.

Equations~(\ref{eq:bias_C})-(\ref{eq:m1}) demonstrate that the fractional bias or error is smaller if the length of the data vector is reduced or the number of simulations is increased. However, they also demonstrate a potential bottleneck when analyzing next-generation surveys. Take, for instance, the DESI survey configuration presented in \cite{DESI_2016}. This survey aims to produce 3D clustering measurements in $0.1$ width redshift bins between $0 < z < 1.9$. In several of these redshift bins, there are three distinct overlapping tracers. It means one might aim to simultaneously fit a joint data vector containing separate but correlated, auto-clustering measurements for each tracer. Furthermore, due to differing observational systematics in the Northern and Southern Galactic Caps, previous surveys such as BOSS \citep{Beutler_2016} and eBOSS \citep{Alam_2021} have often treated the clustering in these two sky-patches as separate, but sometimes correlated, data vectors. Combined, this means we could expect to fit a total data vector consisting of up to 6 correlated auto-clustering statistics in each redshift bin (we will ignore the fact we may also want to measure and fit the cross-clustering between these tracers or sky-patches for now).

Based again on the same previous surveys, we can assume each set of clustering statistics could contain $\sim50$ measurement bins (monopole and quadrupole of the power spectrum between $0\hompc < k < 0.20\hompc$, plus hexadecapole between $0\hompc < k < 0.10\hompc$, all with $0.01\hompc k$-bins). Therefore, the length of the joint data vector for a single redshift bin could be on the order of 300 correlated measurements. Taking a standard EFTofLSS-style approach \citep{d_Amico_2020,Glanville_2022}, we would fit this data vector with a model containing $\sim 6$ cosmological parameters, plus $10$ bias parameters per Galactic Cap, per tracer (so $66$ free parameters in total).  The punchline is that to avoid having to increase our parameter uncertainties by more than $10\%$ ($5\%$) due to covariance matrix noise, we would need $\sim2000$ ($\sim3600)$ simulations for such a redshift bin. Although not impossible, this is clearly computationally demanding and may need to be repeated for other redshift bins or as a result of other systematic tests/validations within the collaboration.

\begin{figure}
    \centering
	\includegraphics[width=1.0\columnwidth]{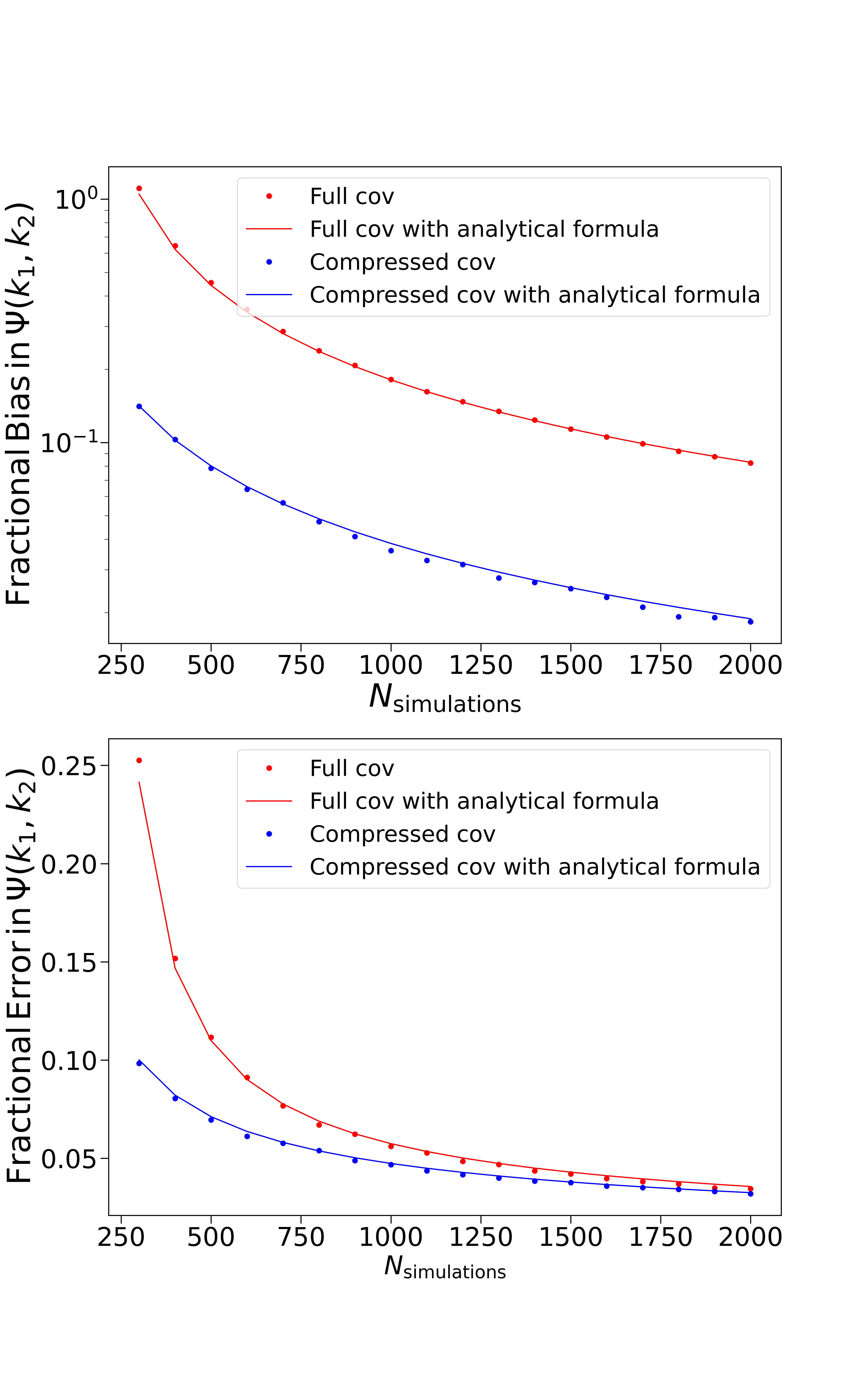}
    \caption{This plot shows the fractional error (bottom) and fractional bias (top) for the precision matrix with data from the BOSS galaxy survey as a function of the number of simulations used in the estimation. In both cases, red points/lines indicate the numerical/analytical results for the full precision matrix, while blue points/lines represent the numerical/analytical results for the compressed precision matrix. Both panels show a clear improvement when using compression.}
    \label{fig:bias_error}
\end{figure}

One way to reduce this burden is to use the MOPED algorithm. We demonstrate this using power spectrum covariance matrices from the BOSS survey (\citealt{Beutler_2021}; detailed more in Section~\ref{sec:data}) estimated using up to 2048 simulations. We estimate the standard deviation and mean of the precision matrix from these simulations with the jackknife method. The numerical result shown in Fig~\ref{fig:bias_error} is in excellent agreement with the predictions from equations~(\ref{eq:bias_C}) and (\ref{eq: frac_Cinv}).
These results also show that the compressed precision matrix has a factor of 10 (2) improvement in the fractional bias (error) compared to its uncompressed counterpart for the same number of simulations, motivating its further exploration in this work.

Beyond compression, one could use analytical covariance matrices such that no simulation is required, and no noise of the form presented above is introduced. Therefore, the analytical covariance matrix can produce tighter constraints than the simulated covariance matrix. However, there are three potential obstacles. Firstly, the analytical covariance matrix has only been shown to work with the BOSS survey \citep{Wadekar_2021}. In this work, we combine BOSS with eBOSS and 6dFGS to create a much larger sample and test whether the covariance matrix can still return unbiased constraints. Secondly, in \citet{Wadekar_2021}, they generate the analytical covariance matrix with the best-fit parameters found using the simulated covariance matrix, which would not necessarily be available. In this work, we will show the impact of using randomly chosen (but with some external prior knowledge) free parameters on the covariance matrix. Furthermore, we also examine the case where the input cosmological parameters for the analytical covariance matrix are more than 3\(\sigma\) away from the mean of the posterior with the simulated covariance matrix. Lastly, \citet{Wadekar_2020} developed the covariance matrix with the Standard Perturbation Theory (SPT) framework, while \textsc{PyBird} was developed using the Effective Field Theory (EFT). Although mapping bias parameters between these two bases is possible, \textsc{PyBird} has additional assumptions such as the ``UV subtraction" in \citet{Perko_2016}, which makes it difficult to find such a relation analytically. In this work, we will show this can be overcome with the local Lagrangian approximation and it has little effect on our final constraints. The end result is a self-consistent framework for analyzing data over multiple redshift bins directly for cosmological parameters without any simulations.

\section{Data}
\label{sec:data}
For tests in this work, we used the combined datasets of the 6-degree Field Galaxy Survey (6dFGS; \citealt{Jones_2009}), the $z1$ and $z2$ redshift bins of the Baryon Oscillation Spectroscopic Survey Data Release 12 (BOSS DR12; \citealt{Alam_2017}) with both NGC and SGC sky-patches, and the LRGpCMASS and QSO (Quasi Stellar Object) samples of the extended Baryon Oscillation Spectroscopic Survey Data Release 16 (eBOSS DR16; \citealt{Alam_2021}) with both NGC (North Galactic Cap) and SGC (South Galactic Cap) sky-patches. The combination of these nine correlated clustering samples creates one of the largest consistent datasets for testing cosmological models; their respective mock catalogs are introduced in \citet{Beutler_2021} and the dataset itself is presented thoroughly in section 5 therein and section 3.1 of \citet{Glanville_2022} and so not repeated here. For all of these data, we account for the window function by convolving the model with a matrix form of the window function, computed as in \cite{Beutler_2021}.

We only analyze the monopole and quadrupole of the power spectrum. The range of wavevectors we have chosen to fit is \(0.01 h \mathrm{Mpc}^{-1} \leq k \leq 0.20 h \mathrm{Mpc}^{-1}\) for all data sets. The result is a single joint data vector consisting of 342 measurement bins. As a ballpark number, from our fiducial fitting method, we find that approximately half of our cosmological information comes from the BOSS data.

\section{Theory}
\label{sec:theory}
This section presents our analytic estimates for the covariance matrices from these datasets, the theoretical basis for applying the MOPED algorithm to the EFTofLSS model, and the implementation of the Taylor expansion method to interpolate the power spectrum. The detailed derivation of the analytical covariance matrix, MOPED compression, and the Taylor expansion method are described in \citet{Wadekar_2020}, \citet{Heavens_2000}, and \citet{Colas_2020} respectively, so they would not be repeated here. We focus on the implementation of these three methods to \textsc{PyBird} in this session.

\subsection{Analytical covariance matrix}
\label{sec:ACM}
The analytical covariance matrix \citep{Wadekar_2020} uses a different bias basis than the one in \textsc{PyBird}. Although it is possible to map bias parameters between these two bases, the \textsc{PyBird} model invokes some extra assumptions such as the ``UV subtraction" \citep{Perko_2016} compared to other EFT approaches. Consequently, the number of bias parameters in \textsc{PyBird} is less than that in the analytical covariance matrix. Therefore, direct mapping is not possible. However, we know ``\(b_1\)" in SPT and EFT both denote the {\it linear} galaxy bias, so we can use the local Lagrangian relations to estimate other bias parameters. For simplicity, we will denote the bias parameters in the analytical covariance matrix as the ``SPT bias parameters" and the bias parameters in \textsc{PyBird} as ``EFT bias parameters".

From \citet{Wadekar_2020}, the analytical covariance matrix can be expressed as:
\begin{align}
    C^{\mathrm{Anal}} &= C_G(P_l) + C_{\mathrm{BC}}(b_1^*, b_2^*, \gamma_2^*) + C_{\mathrm{LA}}(b_1^*, b_2^*, \gamma_2^*) + \nonumber \\ &C_{\mathrm{BC}}(b_1^*, b_2^*, \gamma_2^*) + C_{T_0}(b_1^*, b_2^*, \gamma_2^*, b_3^*, \gamma_3^*, \gamma_2^{x, *}, \gamma_{21}^*) + \nonumber \\
    &C_{\mathrm{SN}}(b_1^*, b_2^*, \gamma_2^*)
    \label{eq:anal_cov}
\end{align}
where the model power spectrum \(P_l\) is given by \citep{d_Amico_2020}
\begin{equation}
    P_l = \sum b_{G, i} P_{l, \mathrm{lin}, i}(\Vec{b}_{\mathrm{NG}}) + P_{l, \mathrm{const}}(\Vec{b}_G) 
    \label{eq:P_model}
\end{equation}
and 
\begin{align}
    \Vec{b}_G &= \{b_3, c_{ct}, c_{r, 1}, c_{r, 2}, c_{\epsilon, 1}, c_{\epsilon, \mathrm{mono}}, c_{\epsilon, \mathrm{quad}}\}, \nonumber \\
    \Vec{b}_{\mathrm{NG}} &= \{b_1, b_2, b_4\}, \nonumber \\
    \Vec{b} &= \Vec{b}_G \cup \Vec{b}_{\mathrm{NG}}. 
    \label{eq:bias1}
\end{align}

Here, we use \(*\) to denote the bias parameters in the bias basis of the analytical covariance matrix, and the ones without \(*\) are in the bias basis of \textsc{PyBird}. We also only highlight the dependence of the analytical covariance matrix (\(C^{\mathrm{Anal}}\)) and the nonlinear model of power spectrum from \textsc{PyBird} (\(P_l\)) on the bias parameters since they are the main focus in this section. They also depend on the cosmological parameters and other properties such as number density and FKP weight. The full expression for the Gaussian covariance (\(C_G\)), Local-Averaging covariance (\(C_{\mathrm{LA}}\)), Beat-Coupling covariance (\(C_{\mathrm{BC}}\)), regular trispectrum (\(C_{T_0}\)), and shot-noise (\(C_{\mathrm{SN}}\)) are given in equation (57), (77), (62), (66), and (91, 92) of \citet{Wadekar_2020} respectively. The expression for \(P_{l, \mathrm{lin}, i}(\Vec{b}_G)\) and \(P_{l, \mathrm{const}}(\Vec{b}_{\mathrm{NG}})\) are 
\begin{equation}
    P_{l, \mathrm{const}} = P_l(k, \Vec{b})|_{\Vec{b}_G \rightarrow 0}
    \label{eq:P_const}
\end{equation}
and 
\begin{equation}
    P_{l, \mathrm{lin}, i} = \frac{\partial P_l(k, \Vec{b})}{\partial b_{G, i}}|_{\Vec{b}_G \rightarrow 0}.
    \label{eq:P_lin_i}
\end{equation}

Equation~(\ref{eq:anal_cov}) demonstrates we need to know \(b^u = \{b_1^*, b_2^*, \gamma_2^*, b_3^*, \gamma_3^*, \gamma_2^{x, *}, \gamma_{21}^*\}\) to calculate the analytical covariance matrix. Since we know \(b_1^* = b_1\), we can use the local Lagrangian relation \citep{Eggemeier_2019} to express the non-local SPT bias parameters in terms of $b_{1}$:
\begin{equation}
    \gamma_2^* = -\frac{2}{7}(b_1 - 1),
    \label{eq:g2}
\end{equation}

\begin{equation}
    \gamma_2^{x, *} = -\frac{2}{7}b_2^*,
    \label{eq:g2x}
\end{equation}

\begin{equation}
    \gamma_{21}^* = -\frac{22}{147}(b_1 - 1),
    \label{eq:g21}
\end{equation}

and 
\begin{equation}
    \gamma_3^* = \frac{11}{63}(b_1 - 1).
    \label{eq:g3}
\end{equation}

For the local SPT bias parameters, we find them with the empirical fitting formulae in \citet{Lazeyras_2016}. Their final expressions are  
\begin{equation}
    b_2^* = 0.412 - 2.143b_1 + 0.929b_1^2 + 0.008b_1^3 - \frac{8}{21}(b_1 -1)
    \label{eq:b2}
\end{equation}
and 
\begin{equation}
    b_3^* = -1.018 + 7.646b_1 - 6.227b_1^2 + 0.912b_1^3 + \frac{796}{1323}(b_1-1) - \frac{8}{7} b_2^*.
    \label{eq:b3}
\end{equation}
Comparing to the equations in \citet{Lazeyras_2016}, equation~(\ref{eq:b2}) has an extra \(-\frac{8}{21}(b_1 - 1)\) and equation~(\ref{eq:b3}) has an extra \(\frac{796}{1323}(b_1-1) - \frac{8}{7} b_2^*\). This is because \citet{Lazeyras_2016} uses a different bias basis than \citet{Wadekar_2020}, and these extra factors convert the local bias parameters into the correct bias basis. Now, all parameters needed to calculate the analytical covariance matrix can be written as a function of \(b_1\). We then substitute them into the formulae in \citet{Wadekar_2020} to find the analytical covariance matrix.

\subsubsection{Application to data}

We test the procedure's suitability described above by constructing individual analytical covariance matrices for each of the nine clustering samples we consider. We then treat the full analytical covariance matrix as a block diagonal matrix with the individual covariance matrix sitting in its corresponding block. We assume there is no correlation between different surveys mainly because of the large cosmological distances and weak overlap between the different samples. However \textit{within} the BOSS survey, there is overlap between the $z1$ and $z2$ bins, which cannot be modelled without extending the analytical covariance matrix work of \citet{Wadekar_2021}. For simplicity, we first test the effect of neglecting these cross-correlation terms entirely instead. 

\begin{figure}
    \centering
	\includegraphics[width=1.0\columnwidth]{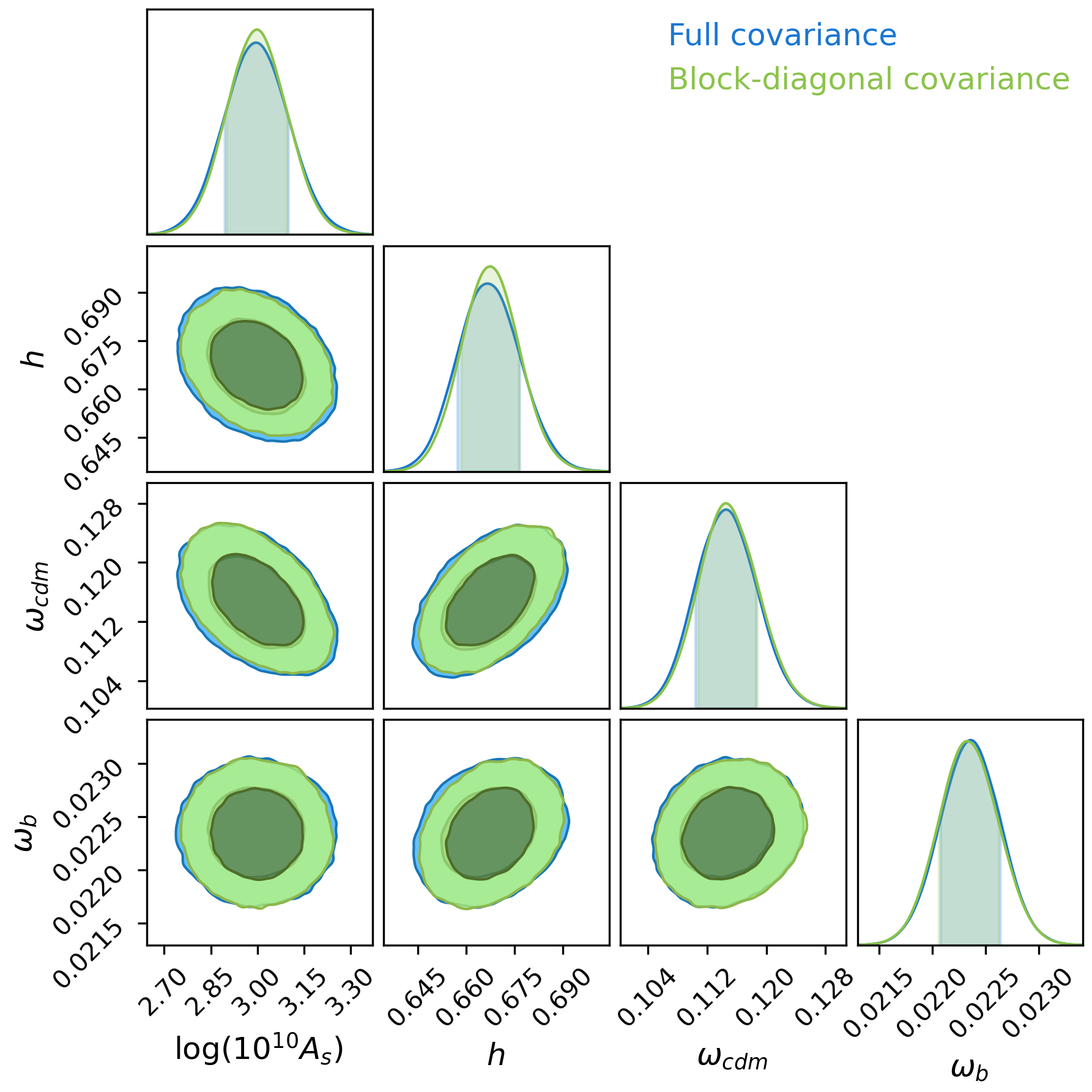}
    \caption{This plot compares the constraints from the Full simulated covariance matrix with the ones from the block-diagonal simulated covariance matrix. Table~\ref{tab:full_vs_block_diag} summarizes the constraints of the cosmological parameters. In general, we see no significant changes in the cosmological parameters, which indicates ignoring the cross-correlation does not affect the mean of the posterior for our data set.}
    \label{fig:full_vs_block_diag}
\end{figure}

We fit our sample with the full simulated covariance matrix and a block-diagonal simulated covariance matrix, where we remove the cross-correlation between the BOSS samples by replacing all non-block-diagonal terms with zero. The priors for our fit are summarized in Table \ref{tab:prior}, and match those in \citet{Glanville_2022} where we fit the same data set with \textsc{PyBird}. The results are summarized in Fig~\ref{fig:full_vs_block_diag} and Table~\ref{tab:full_vs_block_diag}. We found negligible shifts in the mean of the posteriors before and after removing the cross-correlation. There was a small change of <10\% in the parameter uncertainties for $A_{s}$ and $h$, but this difference is tiny. It is also not clear if this is due to sampling noise, because of the absence of physical correlation between these two bins, or because the simulated cross-correlation itself is noisy. As such, we deem it reasonable to simply ignore this cross-correlation component in our fits such that we compare the analytical and simulated covariance matrices on an equal footing.

\begin{table}
    \centering
    \begin{tabular}{c|c}
    \hline
         parameters& prior \\ \hline
         $\ln{(10^{10}A_s)}$& $\mathcal{U}[A_s^{\mathrm{fid}} - 0.8, A_s^{\mathrm{fid}} + 0.8]$\\ \hline
         $h$& $\mathcal{U}[h^{\mathrm{fid}} - 0.08, h^{\mathrm{fid}} + 0.08]$ \\ \hline
         $\Omega_{\mathrm{cdm}}h^2$& $\mathcal{U}[(\Omega_{\mathrm{cdm}}h^2)^{\mathrm{fid}} - 0.04, (\Omega_{\mathrm{cdm}}h^2)^{\mathrm{fid}} + 0.04]$\\ \hline
         $\Omega_bh^2$& $\mathcal{N}[0.02235, 0.00028]$\\ \hline
         $b_1$& $\mathcal{U}[0.0, 3.0]$\\ \hline
         $c_2$& $\mathcal{N}[0.0, 2.0]$\\ \hline
         $c_4$& $0.0$\\ \hline
         $b_3$& $\mathcal{N}[0.0, 2.0]$\\ \hline
         $c_{ct}$& $\mathcal{N}[0.0, 2.0]$ \\ \hline
         $c_{r,1}$ & $\mathcal{N}[0.0, 8.0]$ \\ \hline
         $c_{r,2}$ & $0.0$ \\ \hline
         $c_{\epsilon, 1}$ & $\mathcal{N}[0.0, \frac{\overline{n_g}}{0.025 (h\mathrm{Mpc}^{-1})^3}]$ \\ \hline 
         $c_{\epsilon, \mathrm{mono}}$ & $\mathcal{N}[0.0, 2.0]$ \\ \hline
         $c_{\epsilon, \mathrm{quad}}$ & $\mathcal{N}[0.0, 2.0]$
    \end{tabular}
    \caption{The priors on the cosmological parameters and the nuisance parameters for this work. We follow \citet{d_Amico_2020} to set \(c_2 = \frac{b_2 + b_4}{\sqrt{2}}\) and \(c_4 = \frac{b_2 - b_4}{\sqrt{2}} = 0\) because they found \(b_2\) and \(b_4\) are generally anti-correlated in simulations and observational data. Here, \(\mathcal{U}\)(\(\mathcal{N}\)) denotes the uniform (Gaussian) prior where the first number is the lower bound (mean) and the second number is the upper bound (standard deviation). These priors are used and validated by \citet{Glanville_2022} to fit the same data set. \(A_s^{\mathrm{fid}}, h^{\mathrm{fid}}, \) and \((\Omega_{\mathrm{cdm}}h^2)^{\mathrm{fid}}\) denotes the fiducial cosmological parameters used to calculate the center of the pre-computed grid of power spectrum. The widths of the priors (0.8, 0.08, and 0.04 for \(A_s^{\mathrm{fid}}, h^{\mathrm{fid}}, \) and \((\Omega_{\mathrm{cdm}}h^2)^{\mathrm{fid}}\) respectively) are to ensure cosmological parameters outside the grid won't be evaluated. The fiducial parameter values are in Table~\ref{tab:initial}, but they are changed depending on the exact tests we run.}
    \label{tab:prior}
\end{table}

\begin{table*}
    \centering
    \begin{tabular}{c|c|c|c|c|c|c|c|c}
    \hline
         & $\ln{10^{10}A_s}$ & shift & $h$ & shift & $\Omega_{\mathrm{cdm}}h^2$ & shift & $\Omega_bh^2$ & shift \\
         \hline
         Full& $2.996^{+0.104}_{-0.102}$ & N.A & $0.667^{+0.010}_{-0.010}$ & N.A & $0.1145^{+0.0041}_{-0.0041}$ & N.A & $0.02236^{+0.00028}_{-0.00028}$ & N.A \\
         \hline
         block-diag& $2.997^{+0.098}_{-0.097}$ & $0.01\sigma$ & $0.667^{+0.009}_{-0.009}$ & $0.00\sigma$ & $0.1145^{+0.0044}_{-0.0037}$ & $0.00\sigma$ & $0.02232^{+0.00030}_{-0.00026}$ & $0.14\sigma$ \\
         \hline
    \end{tabular}
    \caption{This table summarizes the shift in the mean posterior after removing the cross-correlation (\(\Omega_bh^2\) is prior dominated). We found no shift in the mean posterior after removing the cross-correlation. It does change the uncertainties of the constraints on the cosmological parameters, but the changes are less than 10\%.}
    \label{tab:full_vs_block_diag}
\end{table*}

Now we know the cross-correlation has negligible impact on the constraints, we can test whether the analytical covariance matrix can return the same constraints as the simulated one. There are two separate steps to calculate the analytical covariance matrix. Firstly, we need the random catalog, FKP weight, and the survey selection function to calculate and save the window kernels and their normalizations (Equation (3) of \citealt{Wadekar_2020}). These variables quantify the survey geometry and are independent of the \textit{true} cosmological parameters. Therefore, we don't have to re-compute them when we change the cosmological parameters. For the second step, we assume a set of cosmological parameters and calculate the linear power spectrum either from \textsc{CLASS} or \textsc{CAMB}.\footnote{The following cosmological parameters are fixed during our analysis, but they are used to calculate the linear power spectrum. We set \(n_s\) (scalar index) \( = 0.9667, \tau\) (optical depth at reionization) \(= 0.066\), and \(M_{\nu}\) (total mass of neutrino) \(= 0.06 \mathrm{eV}\) with a degenerate neutrino hierarchy. Other cosmological parameters are set to their default value for \textsc{CLASS} or \textsc{CAMB}.} Then, we used the saved window functions, their normalization, the linear power spectrum, and the bias parameters to calculate all the analytical covariance components except for \(C_G\). To calculate \(C_G\), we feed the linear power spectrum and the bias parameters to \textsc{PyBird} to calculate the non-linear power spectrum \(P_l\) without the window function. We don't want to include the window function during this calculation because it has already been computed from the random catalog during the first step. Since only the second step depends on the cosmological parameters, we can re-compute the analytical covariance matrix with the new cosmological parameters by feeding it with the new linear power spectrum and non-linear power spectrum without the window function. 

\begin{figure*}
\centering
\begin{subfigure}{0.32\textwidth}
    \includegraphics[width=\textwidth]{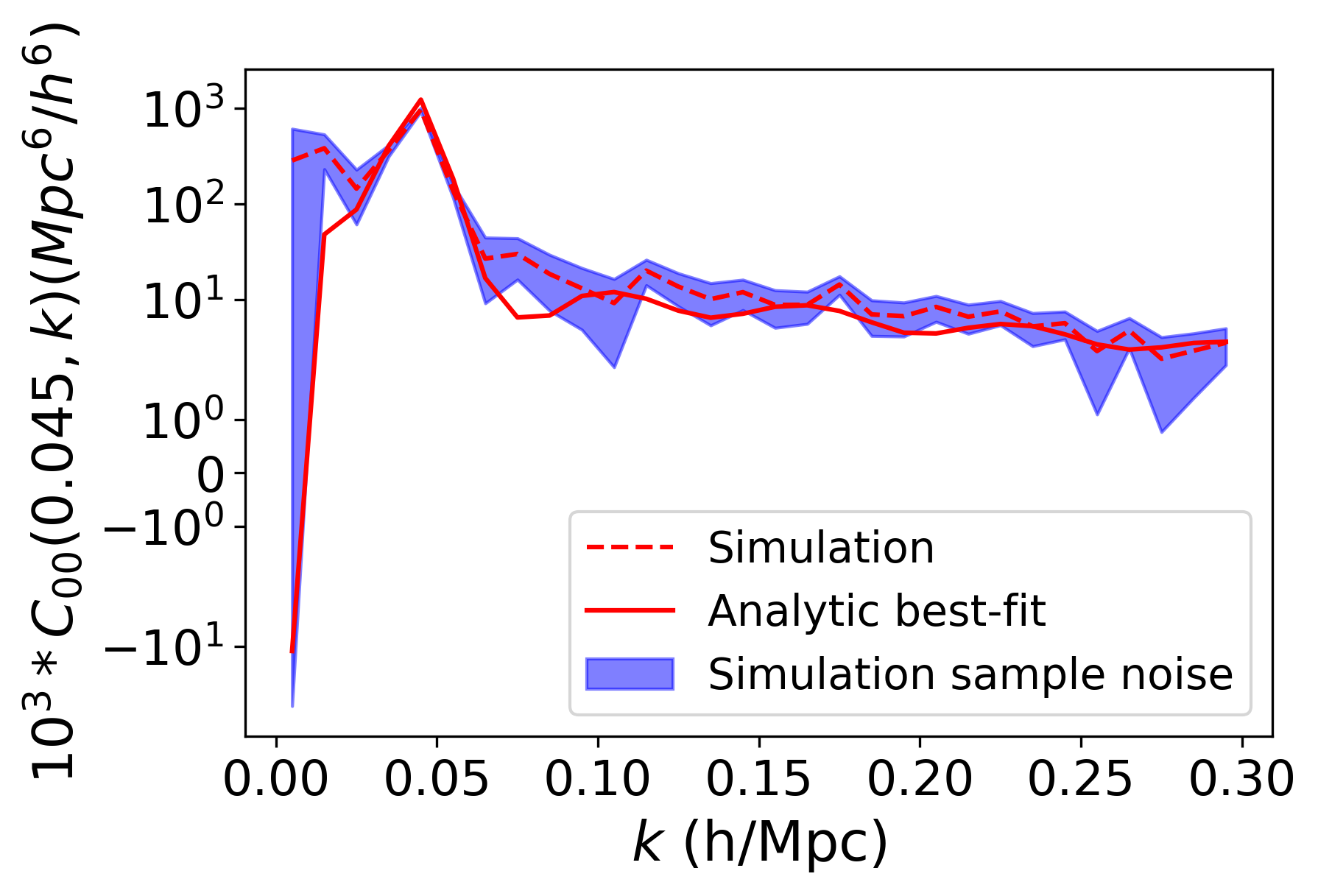}
\end{subfigure}
\hspace{0cm}
\begin{subfigure}{0.32\textwidth}
    \includegraphics[width=\textwidth]{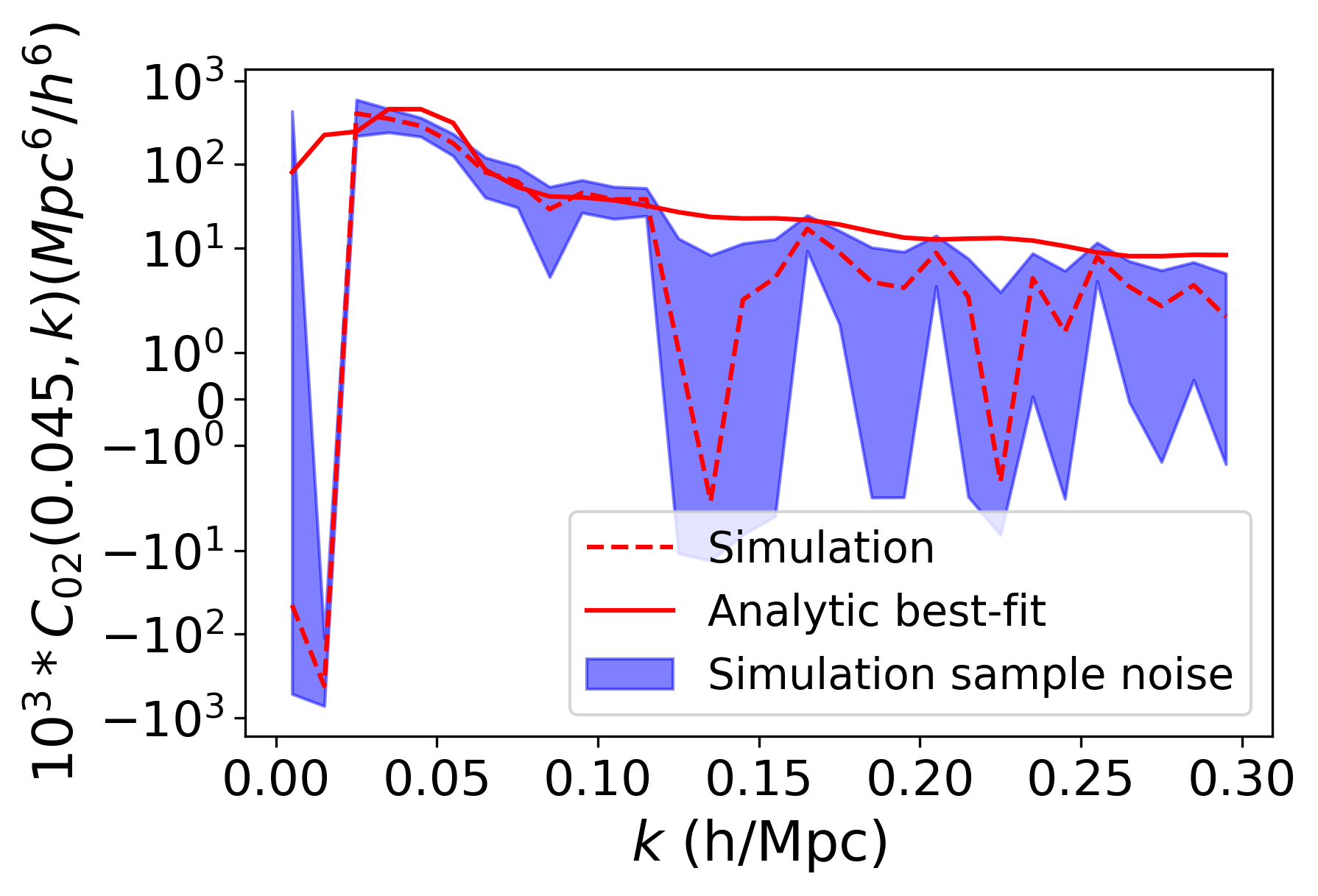}
\end{subfigure}
\hspace{0cm}
\begin{subfigure}{0.32\textwidth}
    \includegraphics[width=\textwidth]{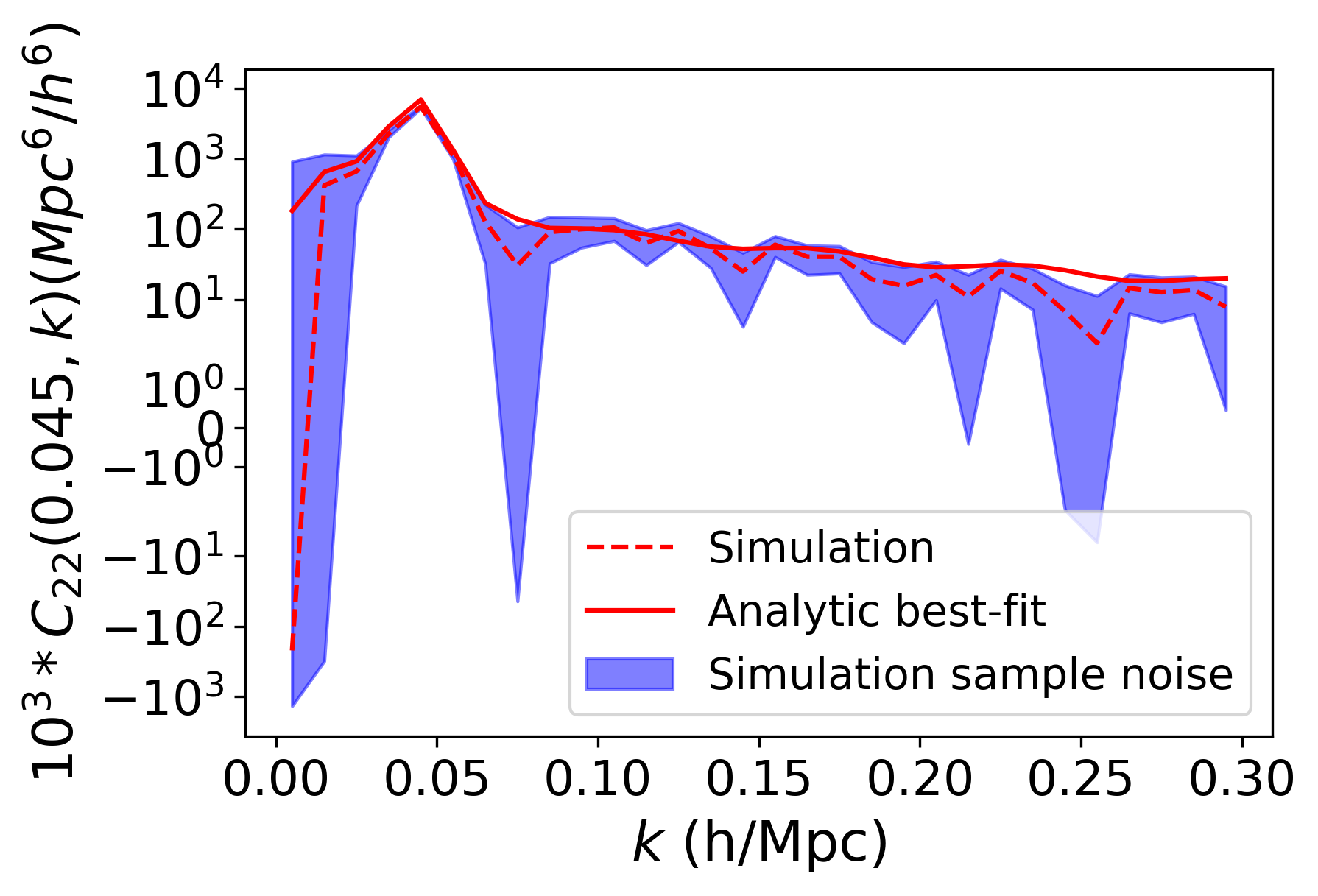}
\end{subfigure}
\hspace{0cm}
\begin{subfigure}{0.32\textwidth}
    \includegraphics[width=\textwidth]{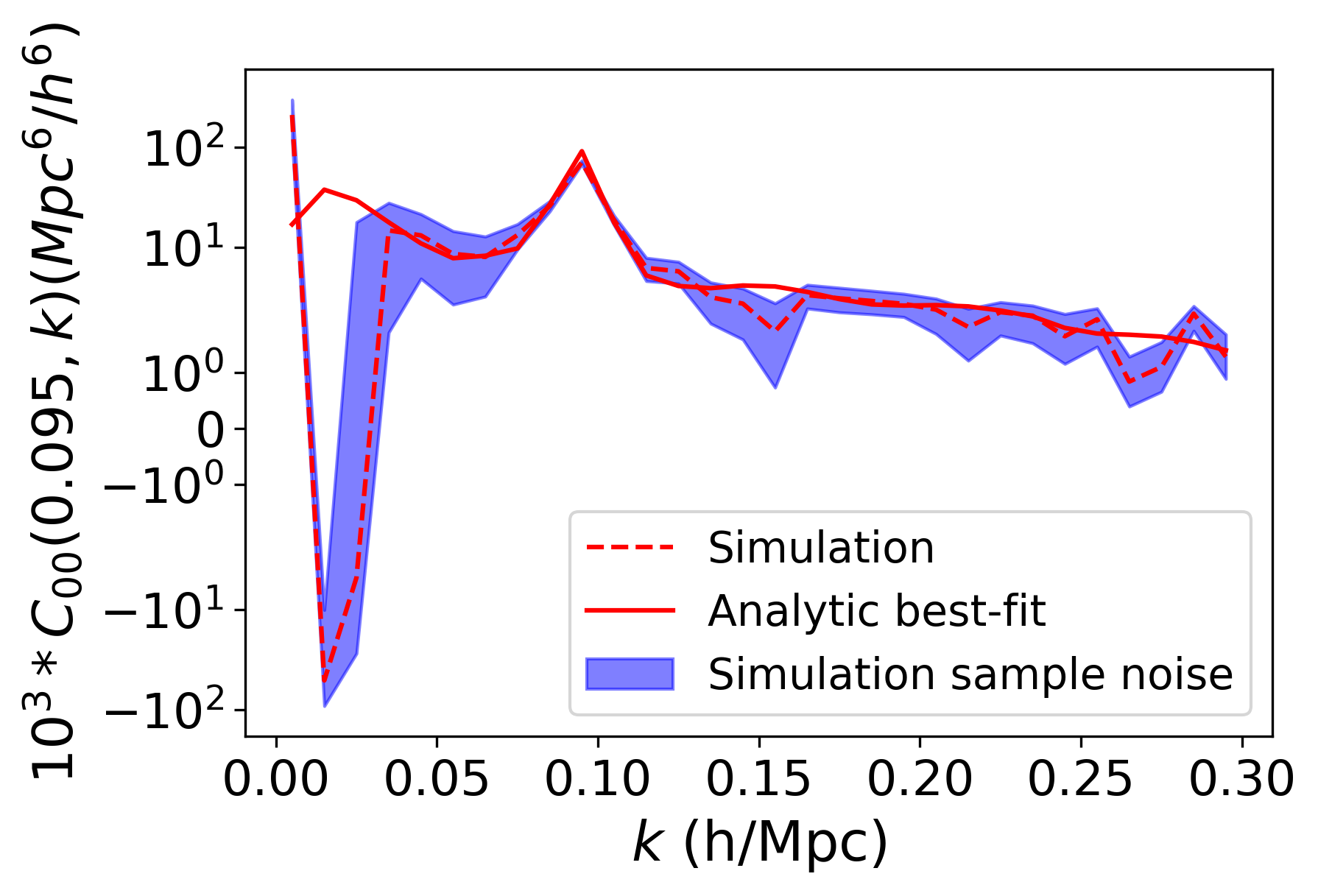}
\end{subfigure}
\hspace{0cm}
\begin{subfigure}{0.32\textwidth}
    \includegraphics[width=\textwidth]{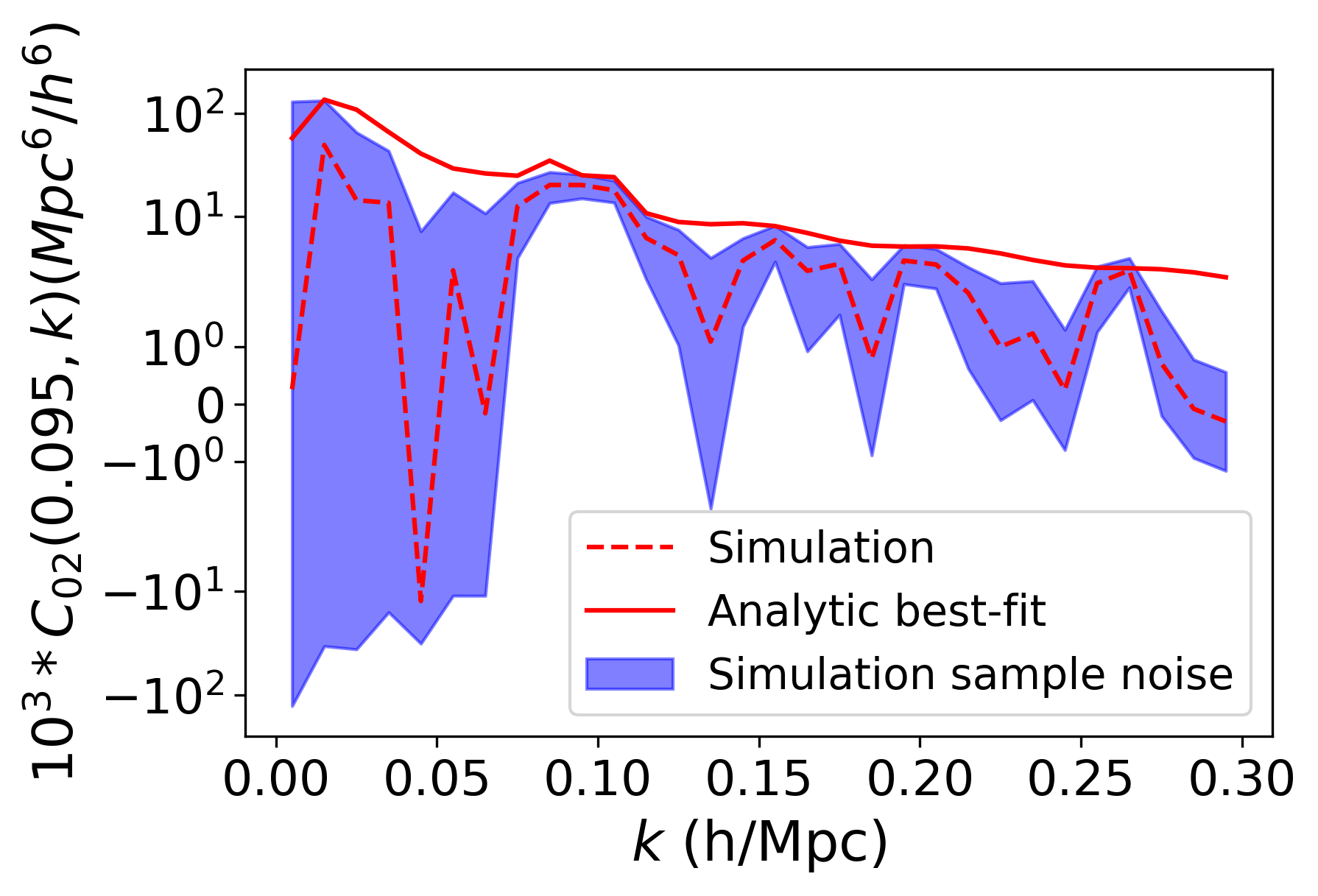}
\end{subfigure}
\hspace{0cm}
\begin{subfigure}{0.32\textwidth}
    \includegraphics[width=\textwidth]{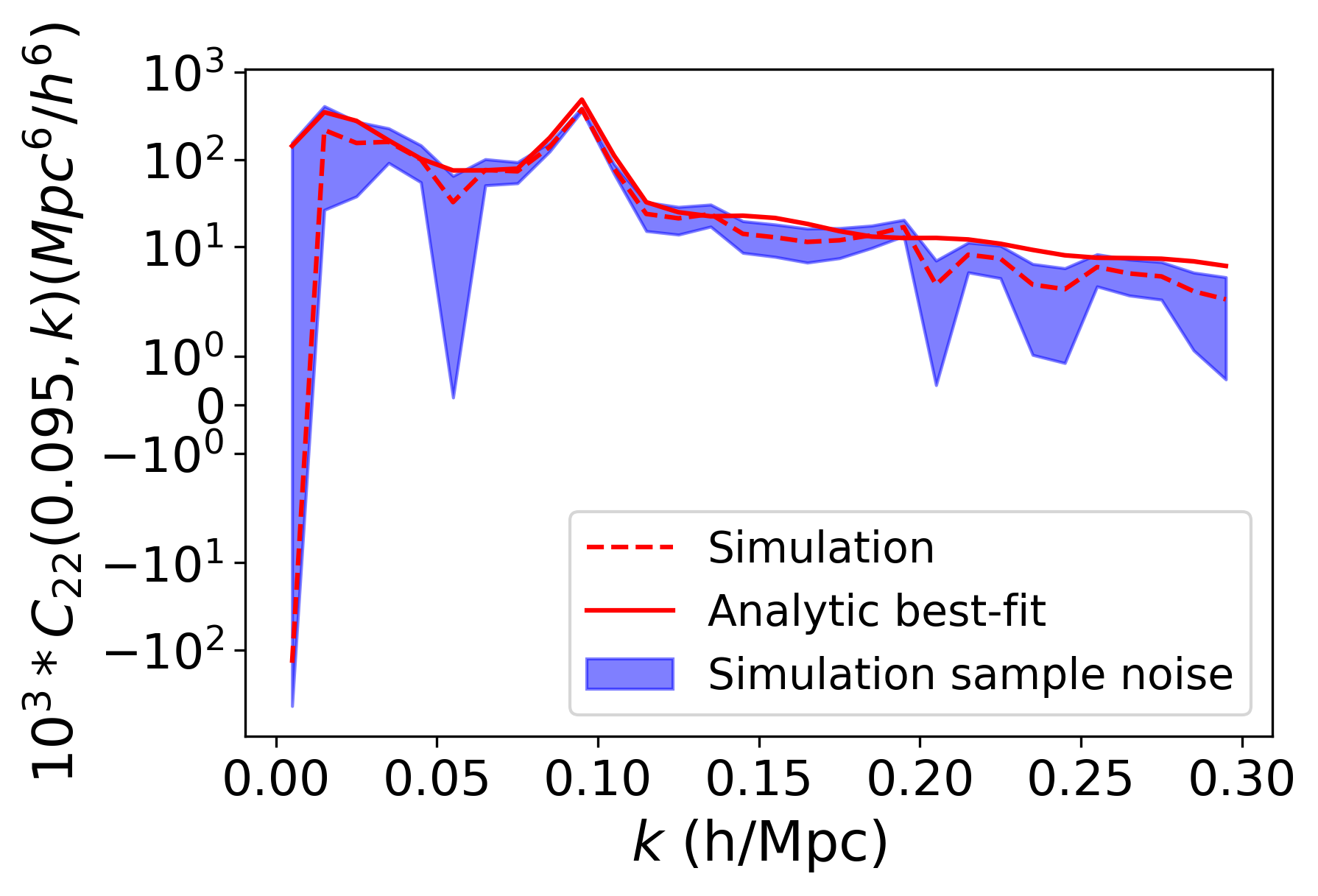}
\end{subfigure}
\hspace{0cm}
\begin{subfigure}{0.32\textwidth}
    \includegraphics[width=\textwidth]{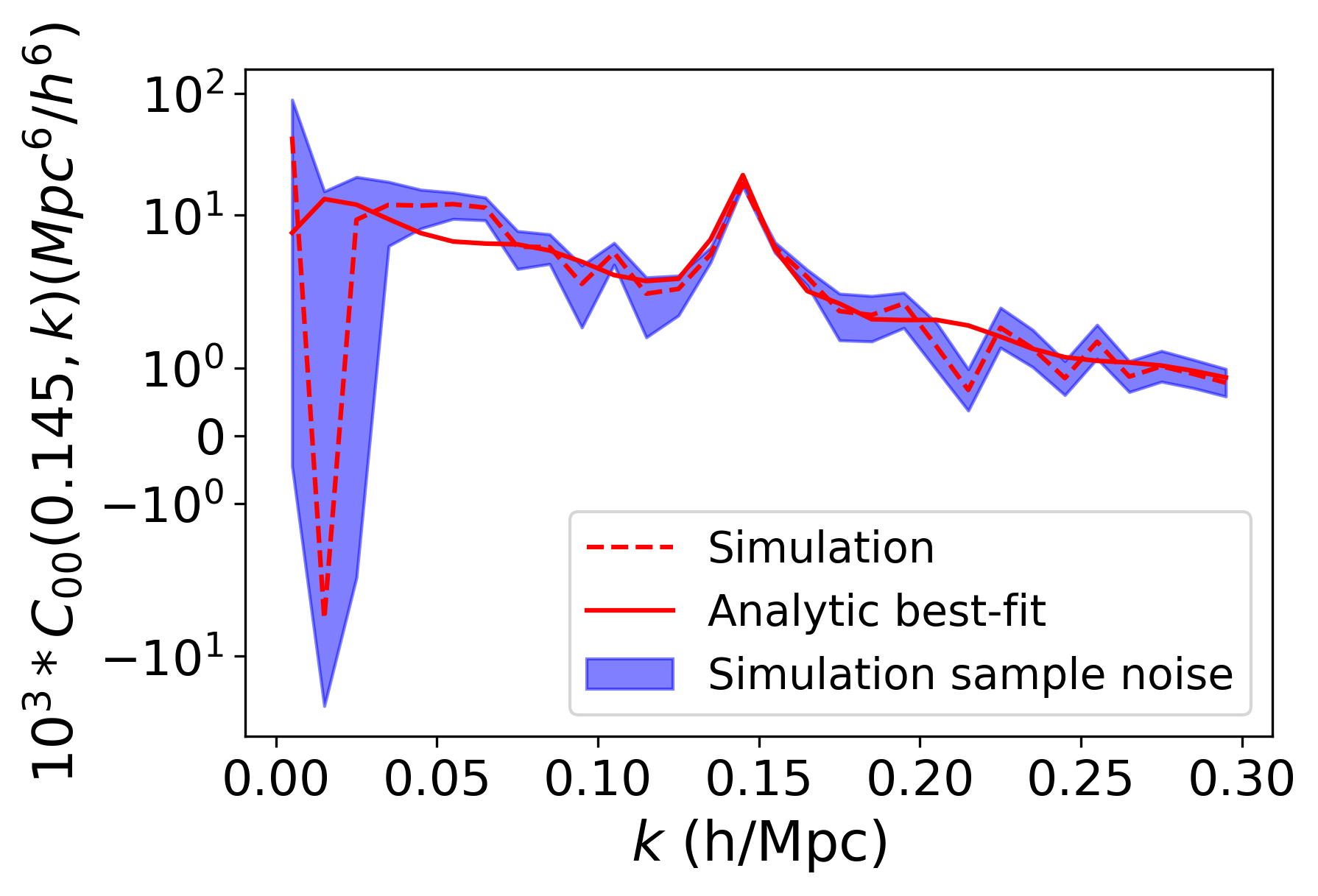}
\end{subfigure}
\hspace{0cm}
\begin{subfigure}{0.32\textwidth}
    \includegraphics[width=\textwidth]{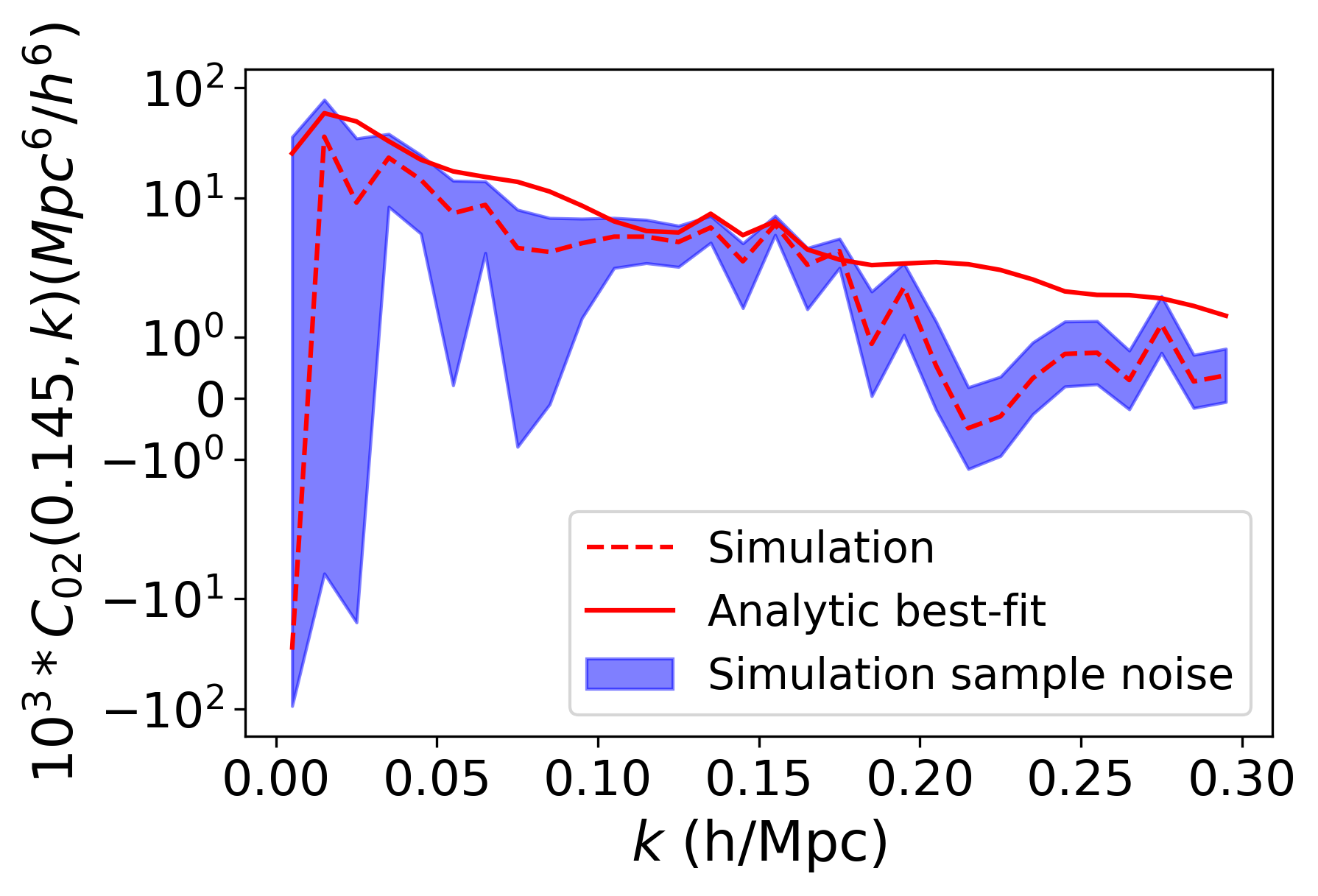}
\end{subfigure}
\hspace{0cm}
\begin{subfigure}{0.32\textwidth}
    \includegraphics[width=\textwidth]{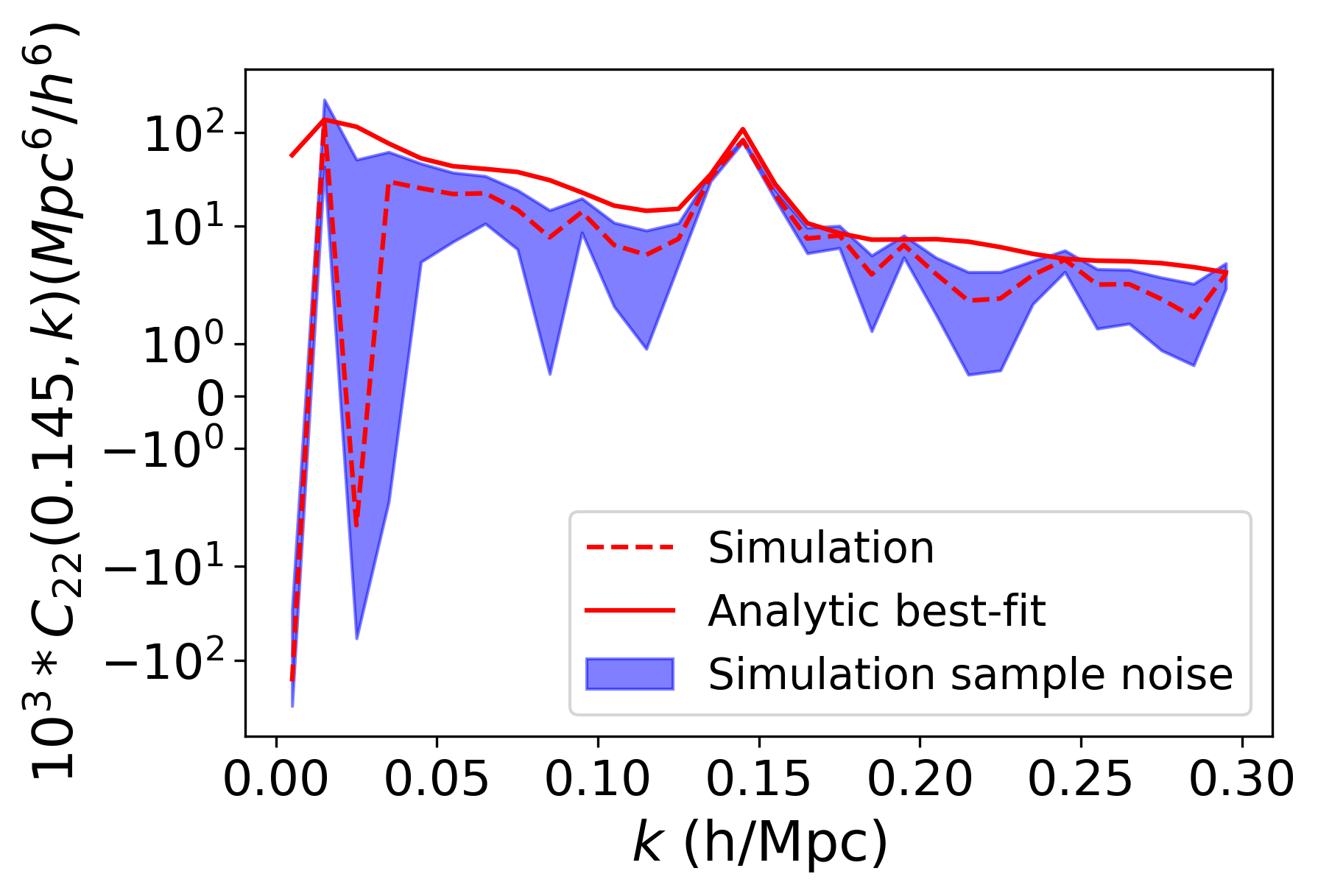}
\end{subfigure}
\hspace{0cm}
\begin{subfigure}{0.32\textwidth}
    \includegraphics[width=\textwidth]{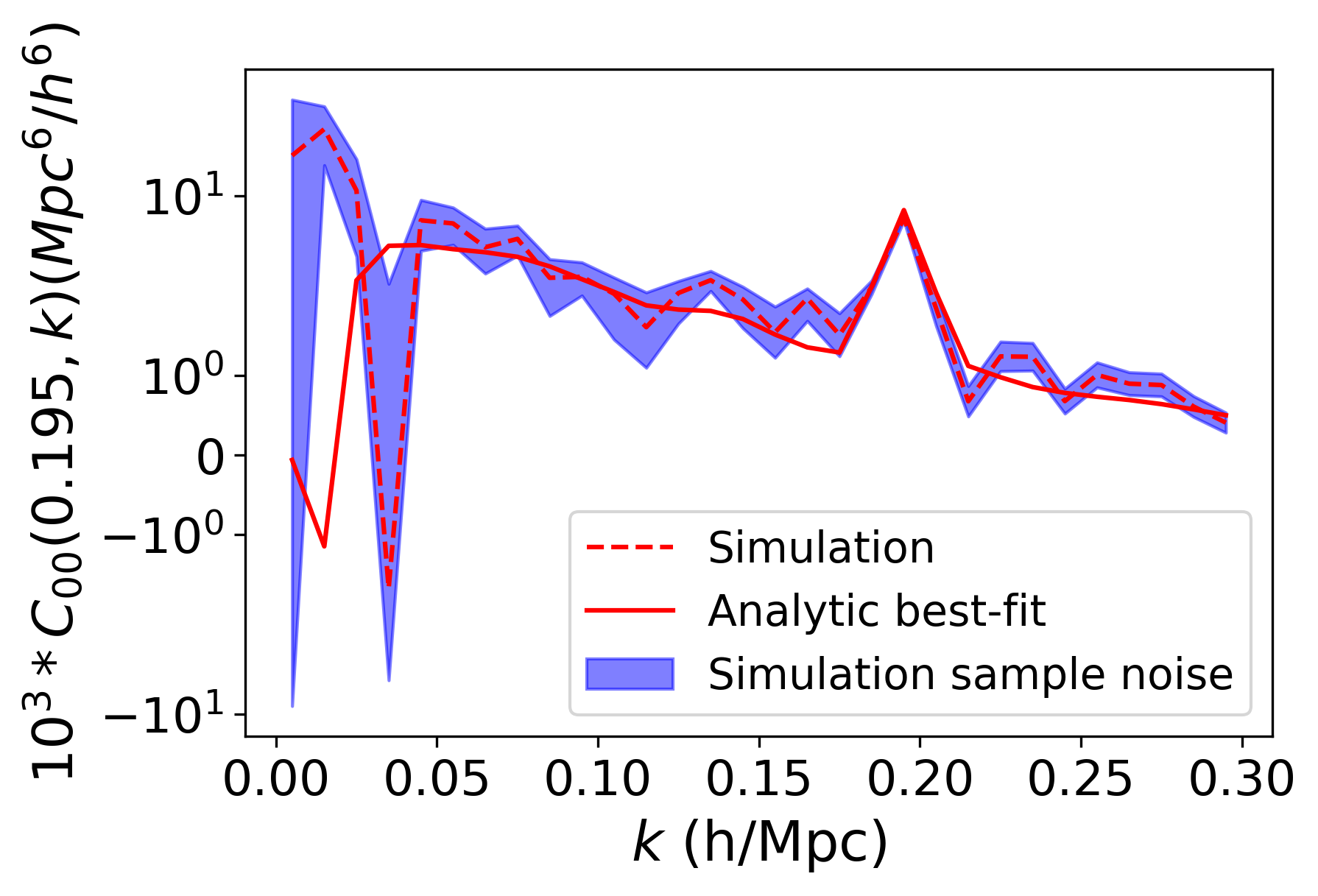}
\end{subfigure}
\hspace{0cm}
\begin{subfigure}{0.32\textwidth}
    \includegraphics[width=\textwidth]{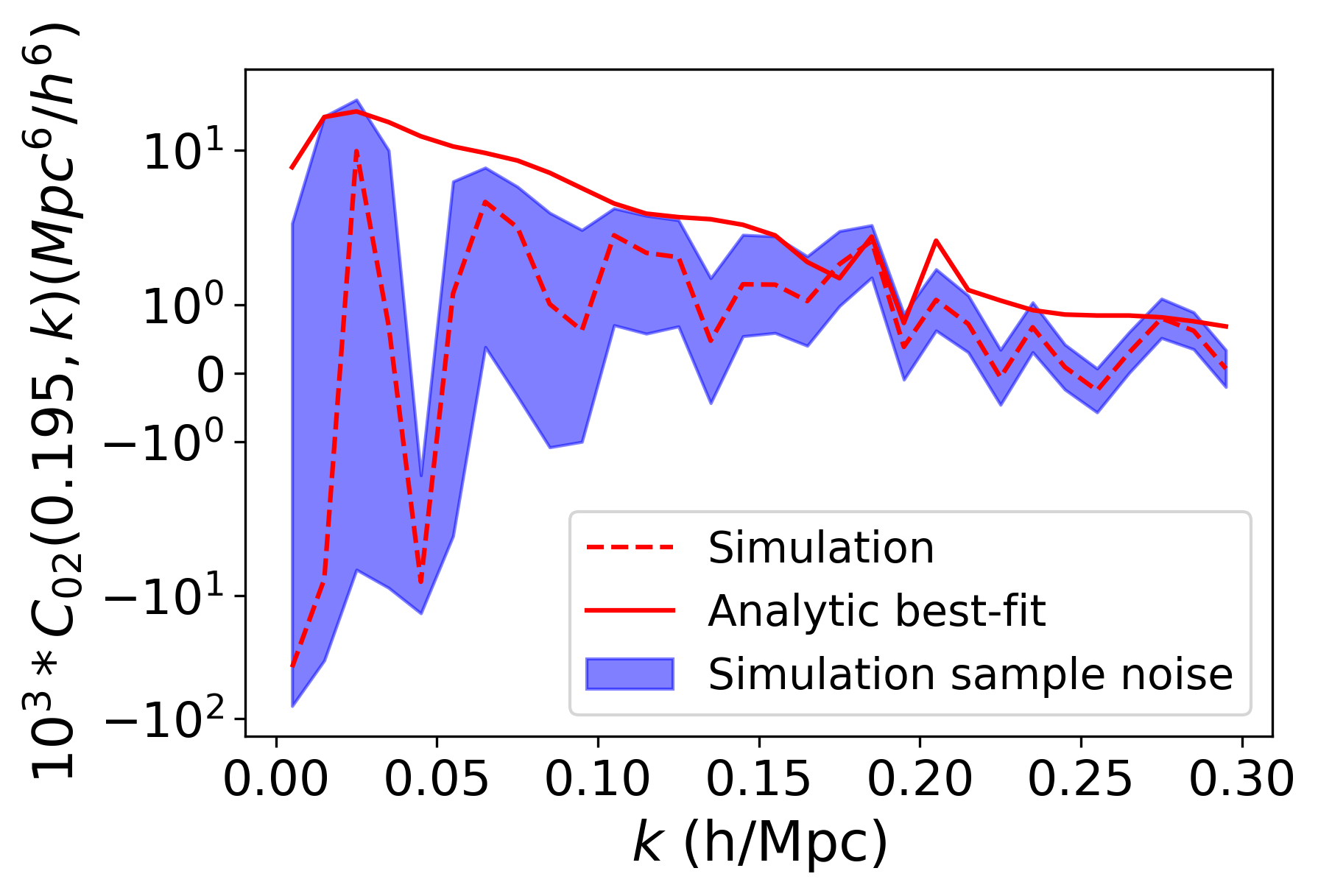}
\end{subfigure}
\hspace{0cm}
\begin{subfigure}{0.32\textwidth}
    \includegraphics[width=\textwidth]{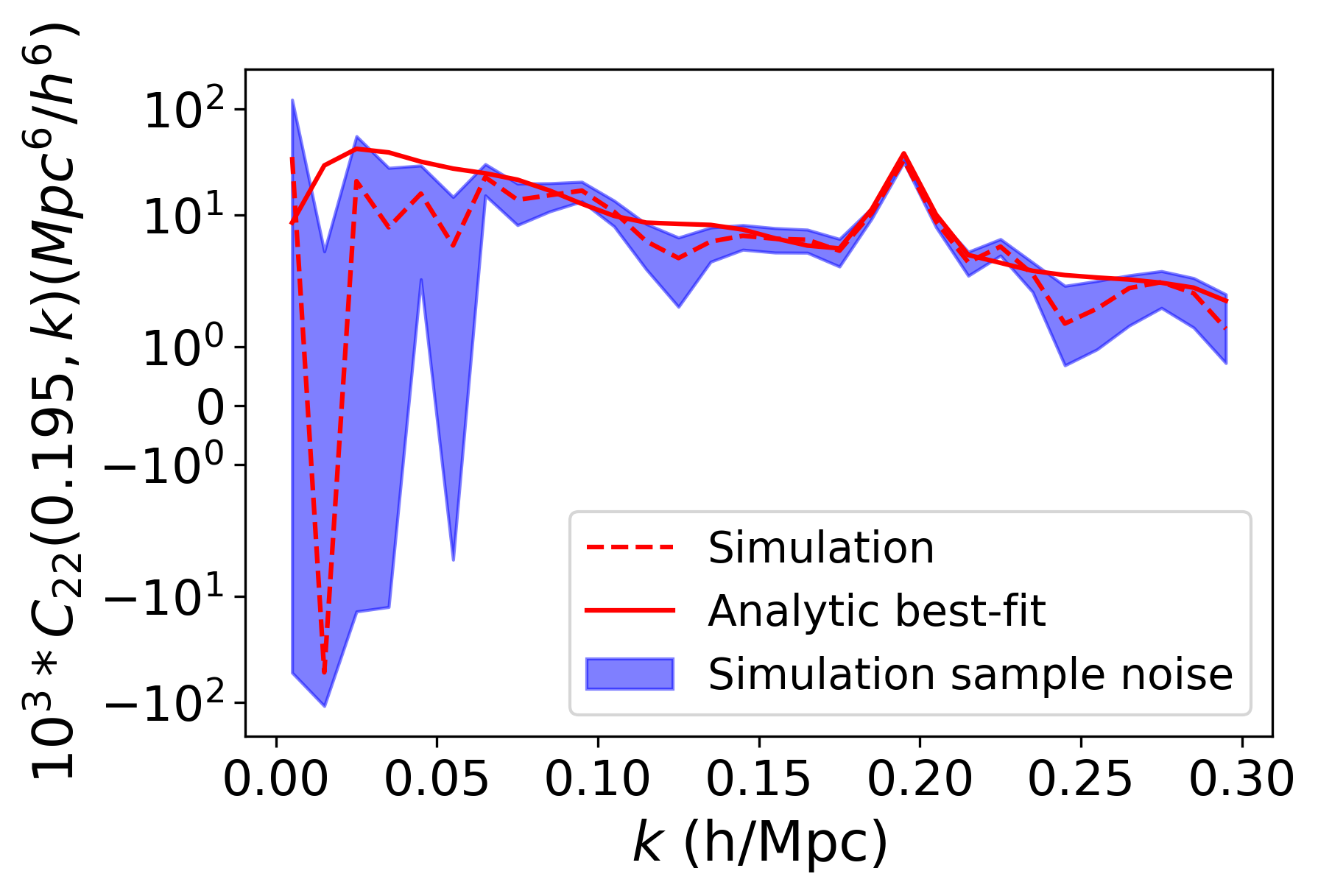}
\end{subfigure}
\caption{This plot compares the analytical covariance matrix with the simulated one for the eBOSS LRG NGC sample. The analytical covariance matrix is generated using the best-fit cosmological and bias parameters from the first iteration, which itself used the ``Template" parameters in Table \ref{tab:initial} and no MOPED compression (A1 in Table \ref{tab:cosmo}). The blue band corresponds to the sampling noise of the simulated covariance matrix. From left to right, we show the auto-covariance matrix for the monopole, the cross-covariance matrix for the monopole and quadrupole, and the auto-covariance matrix for the quadrupole. From top to bottom, we fix the first wavevector of the covariance matrix to \(0.045h \rm{Mpc}^{-1}\), \(0.095h \rm{Mpc}^{-1}\), \(0.145h \rm{Mpc}^{-1}\), and \(0.195h \rm{Mpc}^{-1}\) and vary the second wavevector from \(0.0 h \rm{Mpc}^{-1}\) to \(0.30 h \rm{Mpc}^{-1}\). The plots show that the analytical auto-covariance matrix is generally consistent with the simulated covariance matrix within the sampling noise limit, especially along the diagonal. The small disagreement along the off-diagonal does not affect the cosmological constraints as shown in section \ref{sec:results}.}  
\label{fig:matrix}
\end{figure*}

Fig.~\ref{fig:matrix} compares the analytical covariance matrix with the simulated covariance matrix for the eBOSS LRG NGC subsample. For the analytical covariance matrix, we use the best-fit cosmological parameters obtained from our first fitting iteration which itself used an analytical covariance matrix computed with the ``Template" parameters in Table \ref{tab:initial}, and without the compression (A1 in Table~\ref{tab:cosmo}).\footnote{You can find the comparison of the covariance matrix for the other eight samples here \url{https://github.com/YanxiangL/Analytical-compressed-cov/tree/main/analytic_vs_sim}.} 
The blue band is the sampling noise of the simulated covariance matrix computed with equation~(\ref{eq:delta_C}). Fig~\ref{fig:matrix} illustrates that the analytical covariance matrix is generally consistent with the simulated covariance matrix within the sampling noise, especially along the diagonal. The small disagreement could be because the local Lagrangian relation is not perfect, so some part of the analytical covariance matrix deviates slightly from the simulated one. Nonetheless, we find that the disagreement in the off-diagonal elements of the covariance matrix does not significantly change the constraints on the cosmological parameters, as the amplitude of these components is generally much smaller than either the diagonals of their respective blocks. \citet{Wadekar_2021} similarly demonstrates that the off-diagonal components of the covariance between the monopole and quadrupole have a marginal impact on the constraint of the cosmological parameters. 

\subsection{The MOPED algorithm}
\label{sec:MOPED}
The Massively Optimized Parameter Estimation and Data compression (MOPED) algorithm has been demonstrated to be able to reduce the number of data points without losing information \citep{Heavens_2000, Heavens_2017}. The algorithm achieves this by keeping the Fisher Matrix of the statistic in question with respect to the parameters of interest the same before and after the compression. Section \ref{sec:motivation} demonstrates the compressed precision matrix gives a smaller fractional error and bias than the uncompressed precision matrix. Therefore, the compression could also provide more reliable constraints for the cosmological parameters. We demonstrate its application to the EFTofLSS model and our large-scale structure dataset here.

Suppose we have \(M\) number of free parameters. We can define a compression matrix, $\comp$, consisting of rows $\comp_{m}$ with \(1\leq m \leq M\) and length equal to the number of data points in the uncompressed data vector. Multiplication of the uncompressed data vector by the compression matrix then reduces the size of the new data vector down to length $M$. By enforcing that the Fisher information is conserved through multiplication with $\comp$, the formulation for each row of the compression matrix is given by \citep{Heavens_2000}
\begin{equation}
  \comp_m = \begin{cases}
 \frac{\icov \boldsymbol{P}_{, m}}{\sqrt{\boldsymbol{P}_{, m}^T \icov \boldsymbol{P}_{, m}}} & m=1 \\
 \frac{\icov \boldsymbol{P}_{, m} - \sum_{q = 1}^{m-1} (\boldsymbol{P}^{T}_{, m}\comp_q)\comp_q}{\sqrt{\boldsymbol{P}_{, m}^T \icov \boldsymbol{P}_{, m} - \sum_{q=1}^{m-1} (\boldsymbol{P}_{,m}^T \comp_q)^2}} & 1 < m \le M
\end{cases}
    \label{eq:compression}
\end{equation}
where the model power spectrum is given by \(\boldsymbol{P}\) and \(\boldsymbol{P}_{, m}\) is the partial derivative of the model power spectrum with respect to the \(m^{\mathrm{th}}\) free parameter. The transpose is denoted by \(T\) in equation~(\ref{eq:compression}). 

To compute the compression matrix $\comp$, we need to know the derivative of the model power spectrum with respect to the cosmological and nuisance parameters. For the derivatives with respect to the cosmological parameters, this is done when calculating the Taylor expansion of the power spectrum. A detailed description is given in section \ref{sec:Taylor}. In this section, we mainly focus on the derivatives of the power spectrum with respect to the nuisance parameters. In \textsc{PyBird}, the non-linear power spectrum calculation is separated into 24 different terms (3 linear terms, 12 loop terms, 6 counter terms, and 3 stochastic terms). Each of these terms is independent of the nuisance parameters. More importantly, the final non-linear power spectrum is the linear combination of these 24 different terms multiplied by the appropriate bias parameters 
\begin{equation}
    P_l = \sum_{n} b^i_n b^j_n P^n_l.
    \label{eq:LC}
\end{equation}
The explicit form of \(b^i_n\), \(b^j_n\), and \(P^n_l\) of the 24 different terms are summarized in Appendix \ref{sec:bias}. Therefore, it is straightforward to analytically find the derivative of the nonlinear power spectrum with respect to the nuisance parameters in \textsc{PyBird}.

The MOPED compression changes the covariance matrix and the power spectrum, so we must change the likelihood function accordingly. Before applying the compression, the power spectrum is expected to follow a Gaussian distribution, so the likelihood function \(\mathcal{L}\) is given by
\begin{equation}
        \log \mathcal{L} =  -\frac{1}{2} (\boldsymbol{P}^W - \boldsymbol{P}_d)^T \icov (\boldsymbol{P}^W - \boldsymbol{P}_d),
    \label{eq:likelihood_be}
\end{equation}
where \(\boldsymbol{P}_d\) is the data power spectrum and \(\boldsymbol{P}^W= \win \boldsymbol{P_l}\) is the model power spectrum after convolving with the window function $\win$. In this work, both the analytical and simulated covariance matrix take the survey window function into account. Therefore, to calculate the compression matrix $\comp$, we need to input the derivative of the non-linear power spectrum convolved with the window function (\(\boldsymbol{P}^W\)) with respect to the cosmological and nuisance parameters. Furthermore, this means the compressed windowed non-linear power spectrum is given by the compression matrix \(\comp\) multiplied by the windowed non-linear power spectrum (\(\boldsymbol{P}^W\)). From \citet{Heavens_2000}, the covariance matrix after compression is the identity matrix, so the likelihood after applying MOPED is given by 
\begin{equation}
    \log \mathcal{L} =  -\frac{1}{2} (\comp \boldsymbol{P}^W - \comp \boldsymbol{P_d})^T (\comp \boldsymbol{P}^W - \comp \boldsymbol{P}_d).
    \label{eq:likelihood_af}
\end{equation}

To speed up the MCMC, \textsc{PyBird} analytically marginalizes over \(\Vec{b}_{\mathrm{NG}}\) (see equation (\ref{eq:bias1})). The likelihood function after the analytical marginalization is given by \citep{d_Amico_2020}
\begin{equation}
    \log \mathcal{L} = \frac{1}{2} F_{1, i} F_{2, ij}^{-1} F_{1, j} + F_0 - \frac{1}{2} \ln{|F_2|} 
    \label{eq: Marg_L}
\end{equation}
where 
\begin{align}
    &F_{2, ij} = (P^W_{l,\mathrm{lin}, i})^T \icov P^W_{l,\mathrm{lin}, j} \nonumber \\
    &F_{1, i} =  (P^W_{l, \mathrm{const}})^T \icov P^W_{l,\mathrm{lin}, i} + P_d^T \icov P^W_{l,\mathrm{lin}, i} \nonumber \\
    &F_0 = -\frac{1}{2}(P^W_{l, \mathrm{const}})^T\icov P^W_{l, \mathrm{const}} + (P^W_{l, \mathrm{const}})^T\icov P_d - \frac{1}{2}P_d^T \icov P_d, 
    \label{eq:F_fun}
\end{align}
where \(P^W_{l,\mathrm{lin}, i}\) and \(P^W_{l, \mathrm{const}}\) are \(P_{l,\mathrm{lin}, i}\) and \(P_{l, \mathrm{const}}\) from equation~(\ref{eq:P_lin_i}) and (\ref{eq:P_const}) respectively after convolving with the window function. 

Now, we have the marginalized likelihood function. The next obstacle is applying the MOPED compression to \(P^W_{l,\mathrm{lin}, i}\) and \(P^W_{l, \mathrm{const}}\). Equation~(\ref{eq:P_model}) shows the windowed nonlinear power spectrum \(\boldsymbol{P}^W\) is just a linear combination of \(P^W_{l,\mathrm{lin}, i}\) and \(P^W_{l, \mathrm{const}}\). The bias parameters are just constants and would not affect the matrix multiplication, so we have 
\begin{equation}
    D P^W = \sum_i P^W_{l,\mathrm{lin}, i} + D P^W_{l, \mathrm{const}}.
    \label{eq:power_add_compress}
\end{equation}
This means we can calculate the compression matrix for \(P^W_{l,\mathrm{lin}, i}\) and \(P^W_{l, \mathrm{const}}\) the same way as \(P^W\), by multiplying the compression matrix to it. After some mathematical manipulation, the expression for the marginalized likelihood stays the same, but one needs to change equation~(\ref{eq:F_fun}) to 
\begin{align}
    &F_{2, ij}^{\mathrm{comp}} = (D P^W_{l,\mathrm{lin}, i})^T D P^W_{l,\mathrm{lin}, j} \nonumber \\
    &F_{1, i}^{\mathrm{comp}} =  (D P^W_{l,\mathrm{lin}, i})^T D P^W_{l, \mathrm{const}} + (D P^W_{l,\mathrm{lin}, i})^T D P_d \nonumber \\
    &F_0^{\mathrm{comp}} = -\frac{1}{2} (D P^W_{l, \mathrm{const}})^T D P^W_{l, \mathrm{const}}+ (D P_d)^T D P^W_{l, \mathrm{const}} - \nonumber \\ 
    &\frac{1}{2}(D P_d)^T D P_d.
    \label{eq:F_fun_comp}
\end{align}
There are many ways to calculate the equation~(\ref{eq:F_fun_comp}) by rearranging the order of the matrix/vector multiplications. For example, for \(F_{2, ij}^{\mathrm{comp}}\) we can write it as \((P^W_{l,\mathrm{lin}, i})^T D^T D P^W_{l,\mathrm{lin}, j}\). However, it is not computationally efficient to calculate \(F_{2, ij}^{\mathrm{comp}}\) this way because \(D^T D\) has the same dimension as the uncompressed precision matrix, which will not speed up the code. Having investigated different approaches, we found the most computationally efficient way was to pre-compute \(D P_d\) and \((D P_d)^T D P_d\) because the data vector does not change during MCMC. Then during each step in MCMC, we calculate \(D P^W_{l,\mathrm{lin}, i}\) and \(D P^W_{l, \mathrm{const}}\) first, then use the transpose operation to find \((D P^W_{l,\mathrm{lin}, i})^T\) and \((D P^W_{l, \mathrm{const}})^T\) because the transpose operation is much faster than matrix multiplication. After transposing, both the model and data power spectrum are compressed, so the matrix multiplication is faster. We then substitute the compressed power spectra into equation~(\ref{eq:F_fun_comp}) and equation~(\ref{eq: Marg_L}) to find the marginalized likelihood. 

Compression has several benefits. The MOPED algorithm significantly reduces the dimension of the data vector. Therefore, we will need fewer simulations to estimate the covariance matrix to reach the same precision as shown in Fig \ref{fig:bias_error}. Furthermore, the chi-squared calculation is one of the slowest parts when sampling the posterior with MCMC. Reducing the size of the data vector speeds up the MCMC algorithm. Lastly, we don't have to multiply by the covariance matrix after the compression, which also speeds up the MCMC.

However, there are also caveats --- we need to multiply the compression matrix with the model power spectrum at every step. This could cancel out the potential speed-up. The second caveat is that theoretically, the compression matrix needs to be generated with the best-fit parameters for the compression to be lossless. \citet{Heavens_2000} found that using parameters that are different from the best-fit only marginally increases the credible regions. We demonstrate that the same is true here in the remainder of this work. This is extremely useful in our case because we aim to presume ignorance of the best-fit parameters when generating the analytical covariance matrix and the compression matrix.


\subsubsection{Application to data}
\begin{figure}
    \centering
    \includegraphics[width=0.95\columnwidth]{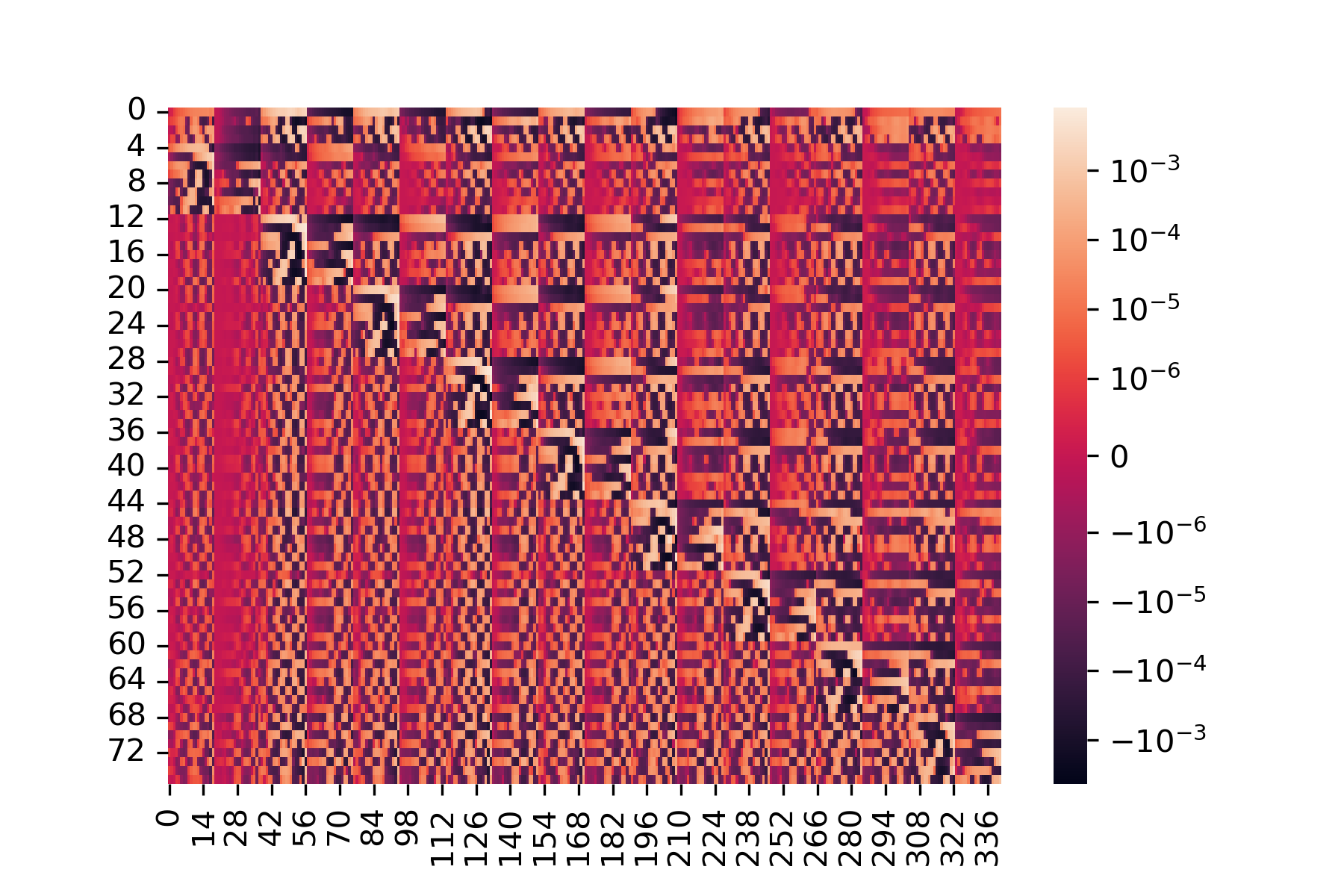}
    
    \centering 
    \includegraphics[width=0.95\columnwidth]{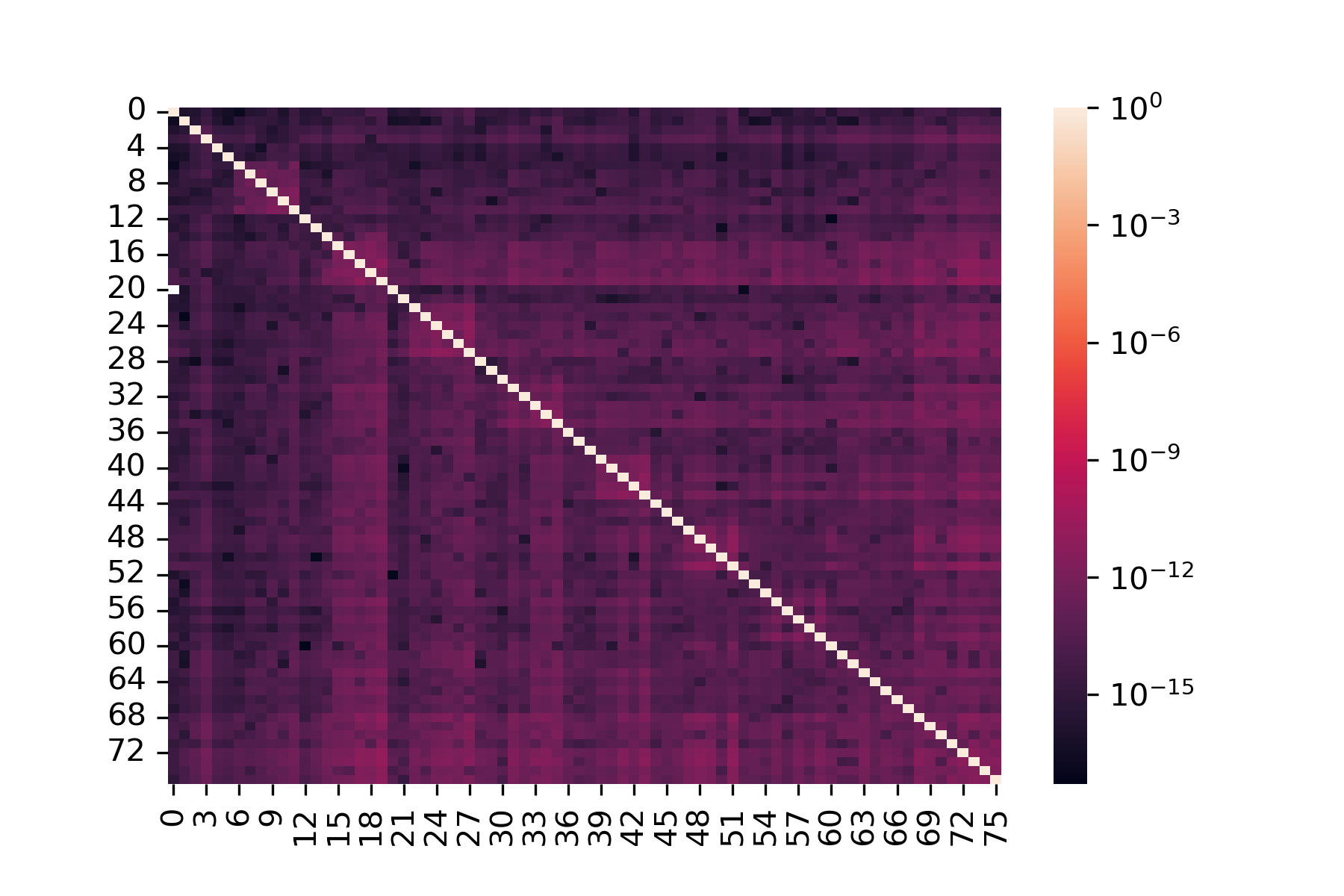}

    \centering
    \includegraphics[width=0.95\columnwidth]{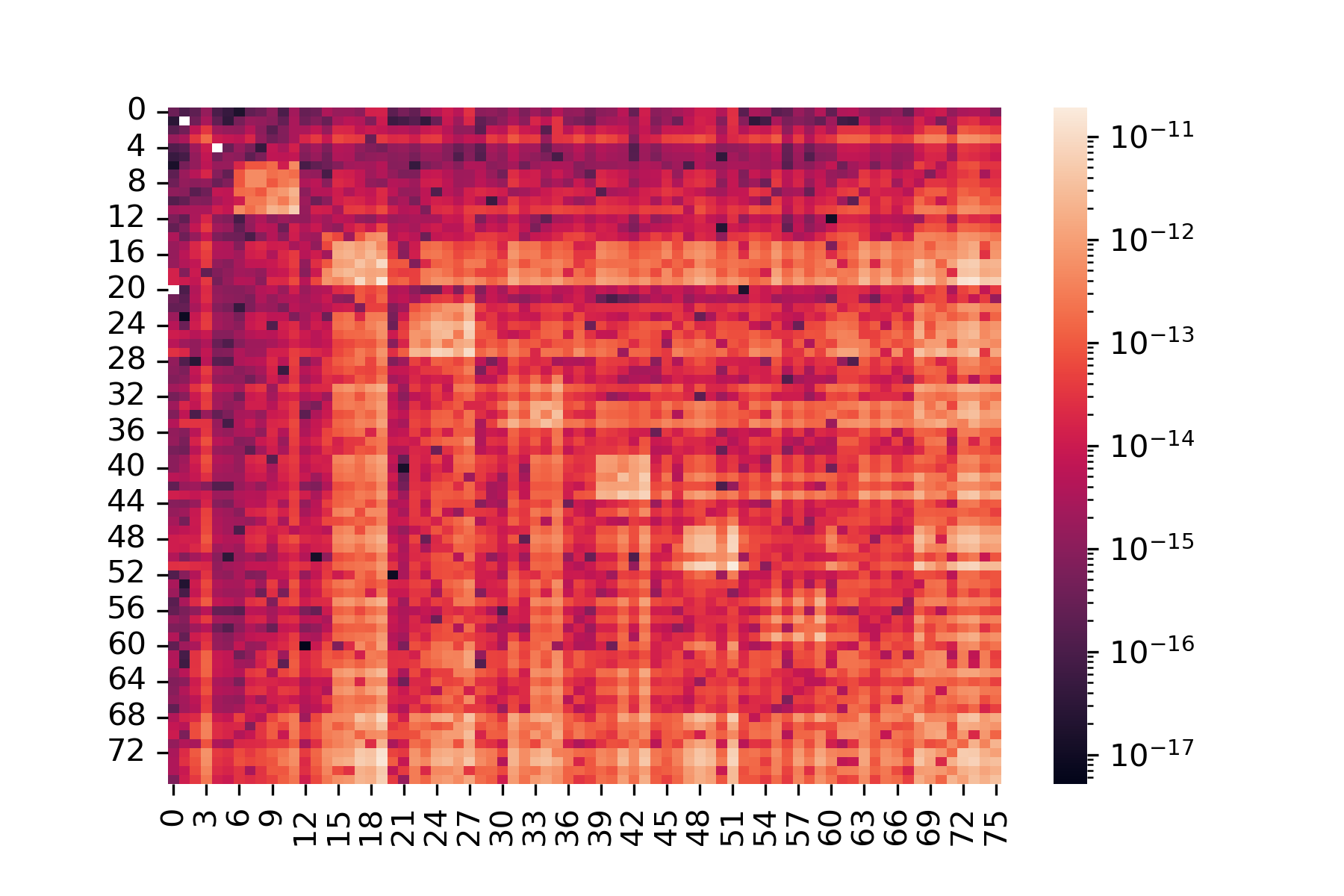}
    \caption{(Top:) A heatmap of the logarithm (base 10) of the compression matrix. Middle: A heatmap of the logarithm (base 10) of the absolute value of the compressed covariance matrix. Except for the diagonal at $\sim\log(1)=0$, all the off-diagonal elements are close to zero. Therefore, the compressed covariance matrix resembles the identity matrix except for some numerical noise. This implies the MOPED compression is successful. Bottom: A heatmap showing the logarithm (base 10) of the absolute difference between the compressed covariance matrix and the identity matrix of the same size. The color bar shows the maximum deviation of the compressed covariance matrix from the identity matrix is on the order of \(10^{-11}\). Again, this could be explained by computational numerical error, so the compression is done successfully.}
    \label{fig:Heatmap_diff}
\end{figure}

As a first test of the MOPED algorithm, we construct the compressed covariance matrix for our full 6dFGS+BOSS+eBOSS data vector and the analytical covariance matrices from Section~\ref{sec:ACM}. In total, we have nine different samples, and each sample has eight different independent free parameters.\footnote{see Table \ref{tab:prior}} We also have four free cosmological parameters shared across all nine samples. After compression, the size of the power spectrum becomes \(8 \times 9 + 4 = 76\). Compared to its original size of 342 measurement bins, the compressed power spectrum is more than four times smaller. 

Lastly, we want to check whether the MOPED compression is successful. Equation~(\ref{eq:likelihood_af}) and \citet{Heavens_2000} show that if the compression is successful, the compressed covariance matrix should be the identity matrix. Fig~\ref{fig:Heatmap_diff} illustrates that the absolute difference between the compressed covariance matrix and the identity matrix is everywhere less than \(10^{-11}\) and much smaller than the amplitude of the compression matrix itself. Hence we consider the differences between the compressed covariance and the identity matrix to be due to the numerical computational error. Therefore, the compression is successful. 

\subsection{Using Taylor expansion to interpolate the power spectrum}
\label{sec:Taylor}
Recent advancements in machine learning allow us to use emulators to compute the power spectrum much faster than the traditional approach. In this work, we use one of the simplest forms of emulators, Taylor expansion, to interpolate the likelihood during the data analysis. This method was developed by \citet{Colas_2020} for \textsc{PyBird}. They found it does not impact the constraints of cosmological parameters with the BOSS DR12 data set. 

It takes three steps to generate the grid of power spectrum for \textsc{PyBird}. Firstly, we want to set the cosmological parameters at the center of the grid. These initial parameters are set according to table \ref{tab:initial}. We have two different sets of tests: "Template" and "Guess". If we try to use the analytical covariance matrix and MOPED compression for future surveys, we will not know the best-fit parameters beforehand. However, we can make educated guesses of the best-fit parameters based on previous surveys such as Planck \citep{Planck_2018}. The values in the ``Template" set are taken from the Planck 2018 constraints. For the second set of parameters, they are chosen such that all cosmological parameters except \(\Omega_bh^2\) are more than \(3\sigma\) from the mean of the posterior with the simulated covariance matrix. We did not do the same thing for \(\Omega_bh^2\) because the BBN priors for the galaxy surveys typically dominate it. Furthermore, we also need the nuisance parameters to calculate the analytical covariance matrix and the compression matrix. For simplicity, we assume \(b_1 = 1.8\) and all other parameters equal to 0.5 for both "Template" and "Guess" sets. During our analysis, we find only \(b_1\) influences the final constraints because it strongly degenerates with \(A_s\). This is expected since both change the amplitude of the power spectrum. Furthermore, we can also get \(b_1\) independently through HOD (Halo Occupation Distribution) analysis. Therefore, we only vary the cosmological parameters, not the nuisance parameters between the two initial conditions. For the second and higher iterations, the input cosmological and nuisance parameters are set to the best-fit of the previous iteration. In this work, the width of each grid cell for cosmological parameters is set to \(\Delta \ln{(10^{10}A_s)} = 0.2, \Delta h = 0.02, \Delta \Omega_{\mathrm{cdm}}h^2 = 0.01,\) and \(\Delta \Omega_bh^2 = 0.0005\). For each parameter, we have nine different grid cells (1 at the center, and four in the positive/negative direction). 

For the second step, we compute the 24 different components of the windowed nonlinear power spectrum (\(P^n_l\) in equation~(\ref{eq:LC})) for each of the grid cells and save them. For the last step, we read in these windowed nonlinear power spectra and use the \textsc{Findiff} package to calculate the first three derivatives of these power spectra with respect to the cosmological parameters. During the data analysis, the derivatives and the grid of windowed nonlinear power spectrum are read in to interpolate the model power spectrum. The first derivative is also used to compute the compression matrix in the equation~(\ref{eq:compression}).

It is worth noting that, from the derivations laid out in Section~\ref{sec:MOPED}, one could alternatively 1) interpolate the un-windowed power spectrum and multiply it by the product of the compression and window function matrices or 2) interpolate the fully compressed model on the grid. We do not implement approach 1) because the derivatives of the windowed power spectrum are required to build the compression matrix, and can be obtained automatically from the Taylor expansion in our adopted approach. Furthermore, the theoretical model vector before convolution is typically evaluated in much narrower bins than the data, such that it would be more expensive to store the grids, and the multiplication of the model with the convolved window would be more computationally expensive. We do not investigate approach 2) in this work because it would require more extensive validation of the Taylor expansion methodology itself (which to our benefit was carried out in \citealt{Colas_2020}), and could be difficult in general as there is no longer the ability to identify a ‘scale’ at which the Taylor expansion may break down. So we leave this to future work.

\begin{table}
    \centering
    \begin{tabular}{c|c|c|c|c}
    \hline 
         & $A_s^{\mathrm{fid}}$ & $h^{\mathrm{fid}}$ & $(\Omega_{\mathrm{cdm}}h^2)^{\mathrm{fid}}$ & $(\Omega_{b}h^2)^{\mathrm{fid}}$\\ \hline
        Template & 3.064 & 0.6774 & 0.1188 & 0.02230\\ \hline
        Guess & 3.350 & 0.7000 & 0.130 & 0.02239\\ \hline
    \end{tabular}
    \caption{These are the initial cosmological parameters to generate the grid of the power spectrum. There are two sets of parameters. The parameters from the "Template" set are not the best-fit parameters but are taken from Planck 2018 \citep{Planck_2018} constraints. The parameters in the "Guess" set are chosen such that the cosmological parameters are more than \(3\sigma\) away from the mean of the posterior with the simulated covariance matrix except for \(\Omega_bh^2\). It is because \(\Omega_bh^2\) is dominated by the BBN prior, so it is unlikely to deviate from its fiducial value.}
    \label{tab:initial}
\end{table}

\section{Result and Discussion}
\label{sec:results}

In this section, we will summarize the result from fitting the 6dFGS, BOSS, and eBOSS power spectra with three different covariance matrices (simulation, analytical, and analytical with compression). Here, we will only show the constraints on the cosmological parameters because they are the most important results. In this work, we use the \textsc{emcee} \citep{ForemanMackey2013} package to analyze the data. We use the default affine invariant sampler for the MCMC and follow \textsc{emcee}'s recommendation that the chain is converged when the number of iterations is more than fifty times longer than the integrated auto-correlation time.\footnote{See \url{https://emcee.readthedocs.io/en/stable/tutorials/autocorr/} for more detail.}

\begin{table*}
    \centering
    \begin{tabular}{@{} c|c|c|c|c|c|c|c|c @{}}
    \hline
         & $\ln{(10^{10}A_s)}$ & shift & 100$h$ & shift & 100$\Omega_{\mathrm{cdm}}h^2$ & shift & 100$\Omega_bh^2$ & shift\\ \hline
        S & $2.997^{+0.098}_{-0.097}(3.121)$ & N.A & $66.7^{+0.9}_{-0.9}(66.5)$ & N.A & $11.45^{+0.44}_{-0.37}(11.31)$ & N.A & $2.232^{+0.030}_{-0.026}(2.235)$ & N.A \\ \hline
        SC & $3.003^{+0.096}_{-0.101}(3.124)$ & $0.06(0.03)\sigma$ & $66.7^{+0.9}_{-0.9}(66.5)$ & $0.00(0.00)\sigma$ & $11.46^{+0.40}_{-0.40}(11.31)$ & $0.02(0.00)\sigma$ & $2.236^{+0.027}_{-0.030}(2.235)$ & $0.14(0.00)\sigma$ \\ \hline
        A1 & $3.021^{+0.096}_{-0.086}(3.137)$ & $0.25(0.16)\sigma$ & $66.7^{+0.9}_{-0.8}(66.5)$ & $0.00(0.00)\sigma$ & $11.45^{+0.39}_{-0.34}(11.30)$ & $0.00(0.02)\sigma$ & $2.236^{+0.027}_{-0.029}(2.236)$ & $0.14(0.04)\sigma$ \\ \hline
        AC1 & $3.030^{+0.090}_{-0.091}(3.131)$ & $0.34(0.10)\sigma$ & $66.7^{+0.9}_{-0.8}(66.5)$ & $0.00(0.00)\sigma$ & $11.48^{+0.35}_{-0.37}(11.34)$ & $0.07(0.07)\sigma$ & $2.234^{+0.029}_{-0.027}(2.236)$ & $0.07(0.04)\sigma$ \\ \hline
        AG1 & $2.695^{+0.030}_{-0.031}(3.084)$ & $3.10(0.37)\sigma$ & $66.8^{+0.9}_{-1.0}(66.8)$ & $0.11(0.33)\sigma$ & $11.39^{+0.27}_{-0.27}(11.31)$ & $0.17(0.00)\sigma$ & $2.234^{+0.028}_{-0.028}(2.236)$ & $0.14(0.04)\sigma$ \\ \hline
        ACG1 & $2.691^{+0.031}_{-0.028}(3.085)$ & $3.14(0.38)\sigma$ & $66.8^{+1.0}_{-0.9}(66.8)$ & $0.11(0.33)\sigma$ & $11.41^{+0.27}_{-0.26}(11.34)$ & $0.10(0.07)\sigma$ & $2.235^{+0.028}_{-0.029}(2.236)$ & $0.11(0.04)\sigma$ \\ \hline
        A2 & $3.005^{+0.098}_{-0.095}(3.115)$ & $0.08(0.06)\sigma$ & $66.5^{+0.9}_{-0.8}(66.4)$ & $0.22(0.11)\sigma$ & $11.46^{+0.38}_{-0.39}(11.27)$ & $0.02(0.10)\sigma$ & $2.234^{+0.030}_{-0.027}(2.236)$ & $0.07(0.04)\sigma$ \\ \hline
        AC2 & $2.999^{+0.094}_{-0.100}(3.127)$ & $0.02(0.06)\sigma$ & $66.7^{+0.9}_{-0.9}(66.5)$ & $0.00(0.00)\sigma$ & $11.48^{+0.39}_{-0.37}(11.35)$ & $0.07(0.10)\sigma$ & $2.236^{+0.028}_{-0.028}(2.236)$ & $0.14(0.04)\sigma$ \\ \hline
        AG2 & $3.012^{+0.087}_{-0.107}(3.125)$ & $0.15(0.04)\sigma$ & $66.5^{+0.9}_{-0.8}(66.3)$ & $0.22(0.22)\sigma$ & $11.42^{+0.42}_{-0.36}(11.28)$ & $0.07(0.05)\sigma$ & $2.236^{+0.028}_{-0.028}(2.236)$ & $0.14(0.04)\sigma$ \\ \hline
        ACG2 & $2.999^{+0.098}_{-0.096}(3.116)$ & $0.02(0.05)\sigma$ & $66.6^{+0.9}_{-0.9}(66.4)$ & $0.11(0.11)\sigma$ & $11.50^{+0.39}_{-0.37}(11.37)$ & $0.12(0.15)\sigma$ & $2.237^{+0.027}_{-0.030}(2.236)$ & $0.18(0.04)\sigma$ \\ \hline
    \end{tabular}
    \caption{The mean of the posteriors of cosmological parameters with different methods. Inside the bracket are the best-fit parameters found with the basin-hopping algorithm in \textsc{Scipy} \citep{Wales_1997}. For the first column, S denotes simulation, C denotes compression, A denotes analytic, 1 denotes the first iteration, and 2 denotes the second iteration. We use two different sets of cosmological parameters to generate the first set of analytical covariance matrix and compression. The first set is with the ``Template" values in Table \ref{tab:initial}, and the second set (denoted by G in the first column) is with the ``Guess" values in Table \ref{tab:initial}, which are $\sim3\sigma$ from the truth. The `shift' columns denotes the difference of each parameter from the ``ground truth'' (row 1) relative to the uncertainties.}
    \label{tab:cosmo}
\end{table*}

\begin{figure*}
    \centering
    \includegraphics[width=0.495\linewidth]{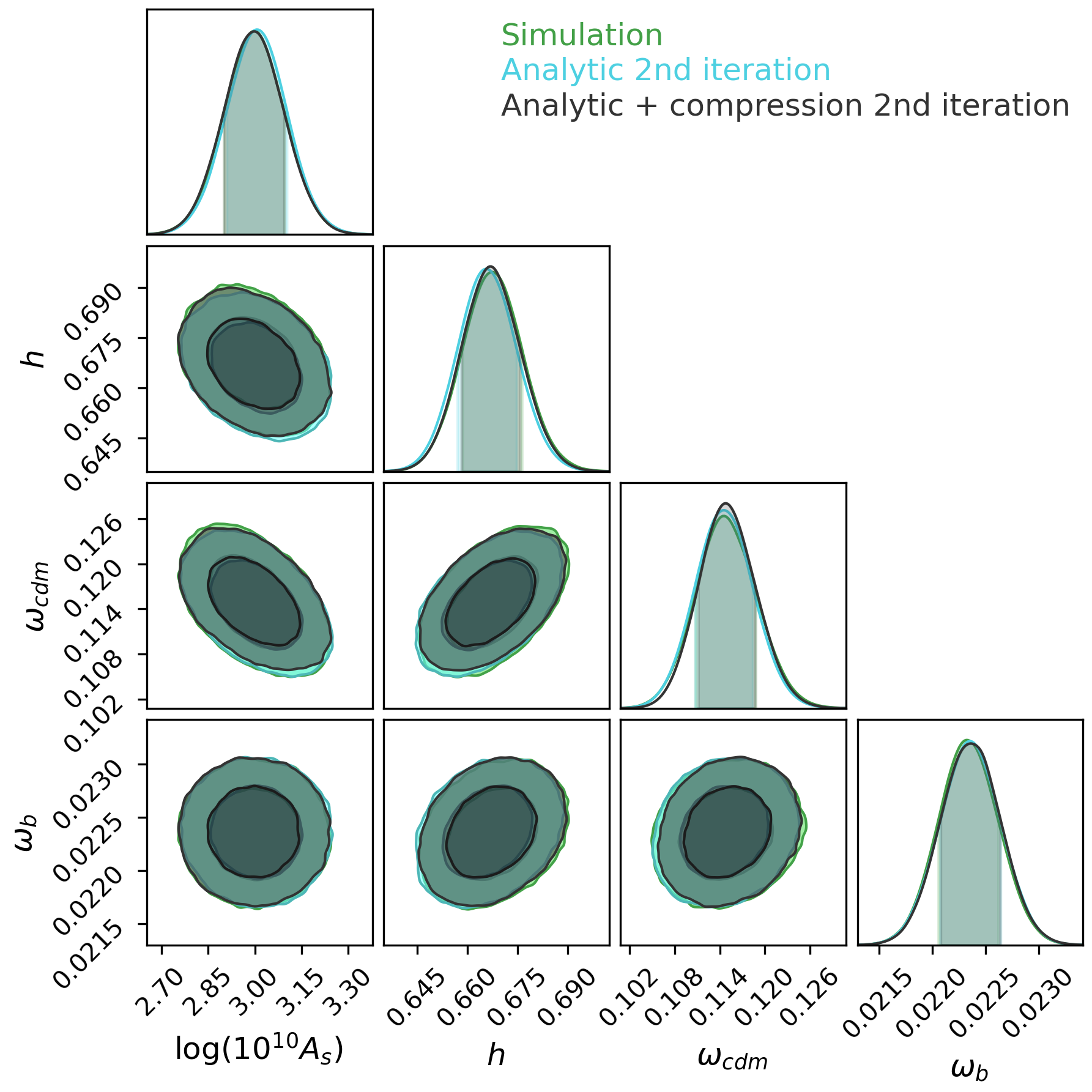}
    \includegraphics[width=0.495\linewidth]{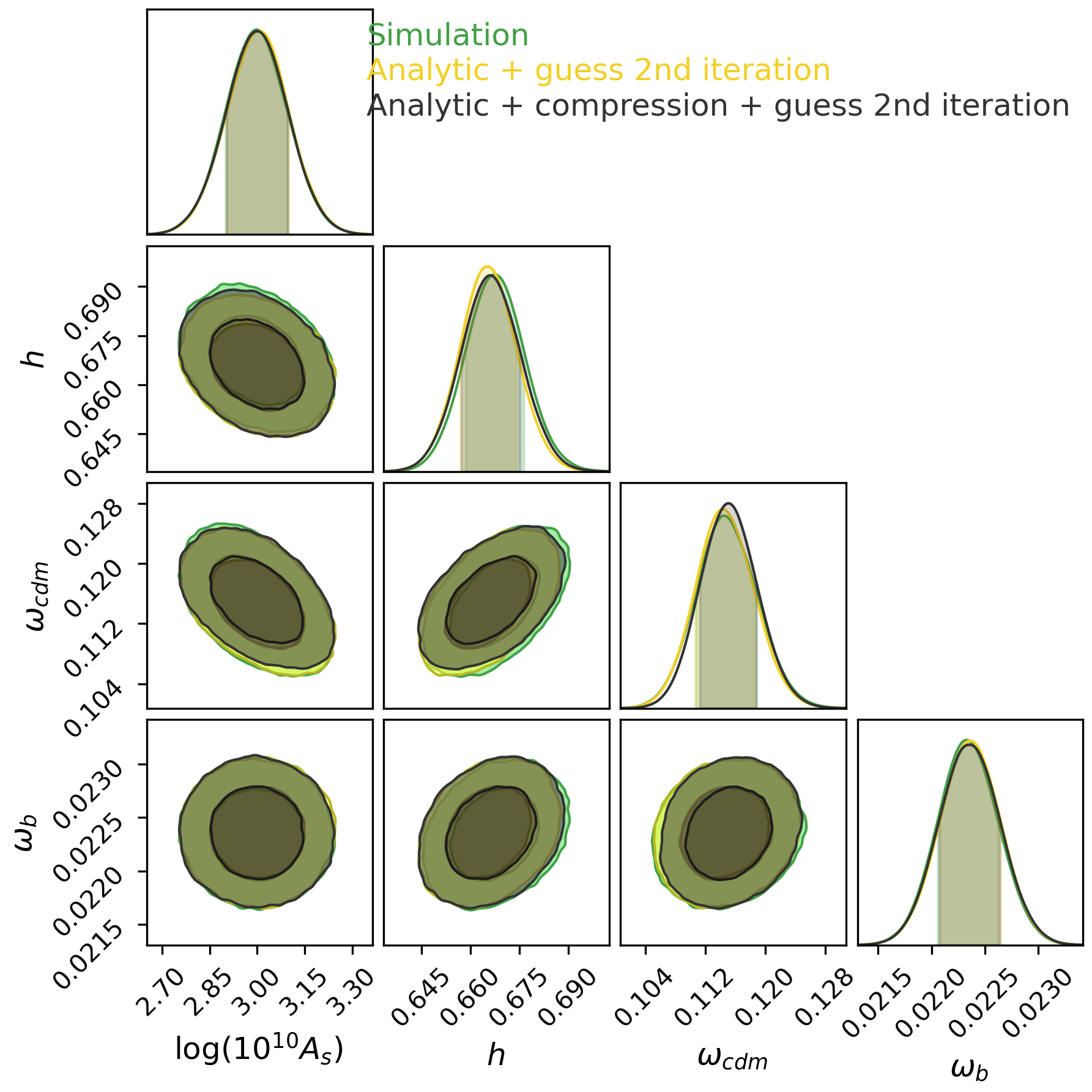}
    \caption{Contour plots for the constraints of cosmological parameters with the ``Template" initial parameters (left) and the ``Guess" initial parameters (right). After two iterations, they are both in excellent agreement with the constraints with the simulated covariance matrix as Table \ref{tab:cosmo} shows.}
    \label{fig:cosmo}
\end{figure*}

The constraints on the cosmological parameters are shown in Fig.~\ref{fig:cosmo} and their respective numerical values in Table~\ref{tab:cosmo}. Inside the brackets are the best-fit values computed with the basin-hopping algorithm from \textsc{Scipy} \citep{Wales_1997}. Both Fig.~\ref{fig:cosmo} and Table~\ref{tab:cosmo} demonstrate the constraints on cosmological parameters using the simulated, analytical, and analytical covariance matrix with compression are in good agreement with one another. For the first column in Table \ref{tab:cosmo}, S denotes simulation, C denotes compression, A denotes analytic, 1 denotes the first iteration, and 2 denotes the second iteration. Additionally, G means we used the ``Guess" parameters in Table \ref{tab:initial} to generate the analytical covariance matrix and the MOPED compression for the first iteration. We treat the constraints with the simulated covariance matrix as the ``ground truth". 

Overall, the MOPED compression introduced \(\lesssim0.1\sigma\) (\(\lesssim0.05\sigma\)) bias to the mean of the posteriors (best-fit) when applied to the simulated covariance matrix. 

Going further and also using an analytical covariance matrix and with the ``Template" initial parameters in Table \ref{tab:initial} to generate the matrix and the MOPED compression, the biases in cosmological parameters are around \(0.3\sigma\) (\(0.1\sigma\)) for the mean of the posterior (best-fit). Furthermore, the cosmological constraints with the analytical covariance matrix from the first iteration are also slightly tighter than the ones with the simulated covariance matrix. 

With the ``Guess" parameters (which are 3$\sigma$ from the truth), we see a much bigger shift, especially for \(\ln{(10^{10}A_s)}\). Its mean of the posteriors is shifted for more than \(3\sigma\) but the best-fit only \(0.4\sigma\). This difference is probably due to the prior volume effect commonly exhibited by EFT models \citep{Carrilho_2023, Simon2023, Holm2023} which is also observed in \citet{Glanville_2022}; generating the covariance matrix with the ``wrong" cosmological parameters seems to significantly worsen this effect. 

In either case, we then perform a second iteration, generating the analytical covariance matrix and the MOPED compression with the best-fit values from the first iteration. After the second iteration, table~\ref{tab:cosmo} shows the biggest deviation of $0.2\sigma$ arises in the constraint on the Hubble parameter with the analytical covariance matrix. Furthermore, if we also apply the MOPED compression, this is reduced to \(0.1\sigma\). This applies regardless of whether we started with the better ``Template'' parameters or the ``Guess'' parameters. The mean of posterior for \(\Omega_bh^2\) is generally shifted by around \(0.14-0.18\sigma\). However, it is worth noting that the constraint on the baryonic density is quite tight because it is dominated by the BBN prior. We applied the BBN prior here because previous analysis shows spectroscopic clustering data has a very weak constraint on \(\Omega_bh^2\). Since the shift in \(\Omega_bh^2\) is much smaller than the width of the BBN prior, We believe this to be a sufficient demonstration of the method.


\begin{table}
    \centering
    \begin{tabular}{c|c|c|c}
    \hline
         & total iteration & time per iteration (s) & total time (h)\\ 
         \hline 
        S & 28815 & $\approx 1.08$ & $\approx 8.64$ \\ \hline 
        SC & 28801 & $\approx 0.25$ & $\approx 2.00$ \\ \hline
        A1 & 32678 & $\approx 1.08$ & $\approx 9.80$ \\ \hline
        AC1 & 31720 & $\approx 0.25$ & $\approx 2.20$ \\ \hline
        AG1 & 121944 & $\approx 1.08$ & $\approx 36.58$ \\ \hline
        ACG1 & 145923 & $\approx 0.25$ & $\approx 10.13$ \\ \hline
        A2 & 34571 & $\approx 1.08$ & $\approx 10.37$ \\ \hline
        AC2 & 33623 & $\approx 0.25$ & $\approx 2.33$ \\ \hline
        AG2 & 35547 & $\approx 1.08$ & $\approx 10.66$ \\ \hline
        ACG2 & 31696 & $\approx 0.25$ & $\approx 2.20$\\ \hline
    \end{tabular}
    \caption{Total time it takes to run MCMC using identical convergence criteria with different methods. The notation of the first column is the same as Table \ref{tab:cosmo}. The bottleneck of MCMC before and after the compression is the chi-squared matrix multiplication. Since all matrices before compression or after the compression have the same size, the time it takes to run each iteration before or after the compression is approximately the same. However, this table shows the covariance matrix has a huge impact on convergence. The analytical covariance matrix generated with the ``Guess" parameters took approximately 4 times more iterations to converge. This is expected since the covariance matrix affects the likelihood which affects the convergence. On the other hand, the total time it takes to run the analysis with the MOPED compression is generally around 4 times faster than without the compression.}
    \label{tab:time}
\end{table}

After two iterations, we found the constraints on cosmological parameters with the ``Template" initial parameters and the ``Guess" initial parameters are consistent with each other despite their initial values being around \(3\sigma\) apart. Therefore, two iterations seem enough to obtain unbiased constraints on cosmological parameters. Table \ref{tab:time} summarizes the time it takes to run each iteration. In general, after the MOPED compression, the MCMC analysis is around four times faster than without the MOPED compression.\footnote{In our work, we first concatenate the data power spectrum and then multiply it by the full covariance matrix to calculate the chi-squared. Alternatively, since we consider our full covariance matrix as block-diagonal one can speed up MCMC by first calculating the chi-squared of each sample individually and then adding it up. In this case, the covariance matrix is much smaller, so MCMC is much faster. We found it only takes around 0.10s per iteration which is faster than the compression method. However, in this case, one can further reduce the size of the matrix by doing compression for each sample individually to still speed up the analysis.} We also found the number of iterations to reach convergence has a significant dependence on the covariance matrix. For the ''Guess" initial parameters which are more than \(3\sigma\) away, they took more than four times more iterations to converge. However, this could be resolved in the future by only running optimization for the first iteration since we are only interested in the best-fit parameters. We used the basin-hopping algorithm from \textsc{Scipy} \citep{Wales_1997} to find the best-fit parameters and it took around one hour to best-fit for AG1 and AGC1. This is more than 10 times faster than the MCMC.

\section{Conclusions}
\label{sec:conclusion}
In this work, we combine the analytical covariance matrix, the MOPED compression, and the Taylor expansion interpolation of the power spectrum for the first time to analyze power spectrum data from the 6dFGS, BOSS, and eBOSS surveys. We demonstrate that this removes the necessity to generate large ensembles of simulations when analyzing such data (a potential bottleneck for future surveys such as the Dark Energy Spectroscopic Instrument (DESI; \citealt{DESI_2019}) and Euclid \citep{Euclid_2011}). while also offering potentially significant speed-ups in model fitting.

For the state-of-the-art dataset tested herein and without assuming \textit{a priori} knowledge of the best-fit parameters, we use two different sets of initial parameters to generate the analytical covariance matrix and the MOPED compression. The ``Template" parameters (within \(1\sigma\) of the mean of the posterior with the simulated covariance matrix) and the ``Guess" parameters (more than \(3\sigma\) from the mean of the posterior with the simulated covariance matrix). The analytical covariance matrix takes around a day to compute, which is much faster than a few months for the simulated covariance matrix. Furthermore, the main bottleneck of the covariance matrix calculation is calculating the window kernels and their normalization in \citet{Wadekar_2020} from the random file, which is independent of the cosmology. The cosmology-dependent part of the calculation that needs to be recalculated for the second iteration only takes around half an hour. After two iterations, we found the mean of the posteriors and best-fit values of cosmological parameters from these two different sets agree with each other and are consistent with the ones with the simulated covariance matrix within \(0.1-0.2\sigma\) regardless of the choice of initial guess. The remaining difference is likely due to the sampling noise. 

We noticed the mean of the posterior of cosmological parameters does not necessarily agree with their best-fit values because of the prior volume effect. The best-fit values should be used to generate the analytical covariance matrix and the MOPED compression for the second iteration. Furthermore, the MOPED compression is able to speed up the MCMC by around a factor of four. This is because the matrix multiplication becomes the main bottleneck with the Taylor expansion. The MOPED compression can significantly reduce the size of the matrices which speeds up the MCMC. 

Further work could investigate whether the compressed data vector itself could be emulated, further decreasing the run time. One could also determine a more rigorous set of convergence criteria for determining how many global iterations to perform, beyond the two we use here, that are suitable for guaranteeing the robustness of results from next-generation surveys

Overall, further investigation of the procedure we have developed here for full-shape fitting offers great potential for significantly reducing the computational cost of directly extracting cosmology from future surveys.


\section*{Acknowledgements}
This research was supported by the Australian Government through the Australian Research Council’s Laureate Fellowship funding scheme (project FL180100168). We like to thank Digvijay Wadekar for his helpful advice on his analytical covariance matrix code. We also like to thank Aaron Glanville for his advice on using \textsc{MontePython} with \textsc{Pybird}. YL is the recipient of the Graduate School Scholarship of The University of Queensland. This research has made use of NASA's Astrophysics Data System Bibliographic Services and the \texttt{astro-ph} pre-print archive at \url{https://arxiv.org/}, the {\sc matplotlib} plotting library \citep{Hunter2007}, CLASS \citep{Blas_2011}, findiff \citep{findiff_2018}, the {\sc chainconsumer} and {\sc emcee} packages \citep{Hinton2016, ForemanMackey2013}. The computation was performed on the Getafix supercomputer and the Tinaroo supercomputer at the University of Queensland. 

\section*{Data Availability}
The code we used to calculate the analytical covariance matrix and the MOPED compression can be found here \url{https://github.com/YanxiangL/Analytical-compressed-cov/tree/main}. To calculate the derivative of the model power spectrum with respect to the cosmological parameters with \textsc{PyBird}, you need to use code in \url{https://github.com/CullanHowlett/pybird/tree/desi}. We use the MontePython code in \url{https://github.com/brinckmann/montepython_public}. The links to the random and data power spectrum can be found in \citet{Jones_2009} for 6dFGS, \citet{Alam_2017} for BOSS, and \citet{Alam_2021} for eBOSS. We also refer the readers to \url{https://github.com/JayWadekar/CovaPT} for the original analytical covariance matrix code. The re-scaled factor applied to the power spectrum can be found in \citet{Beutler_2021}.



\bibliographystyle{mnras}
\bibliography{example} 


\appendix

\section{Appendix A: bias parameters}
\label{sec:bias}
Equation~(\ref{eq:LC}) shows the model power spectrum can be expressed as the linear combination of 24 different terms. These 24 terms are summarized in Table \ref{tab:term_24}. The expressions for \(P11l\) and \(P\mathrm{loop}l\) are very long, so I didn't include them here. The exact expressions can be found in the \textsc{PyBird} Github repository. The linear power spectrum from \textsc{CLASS} is denoted by \(P_{\mathrm{lin}}\) in Table \ref{tab:term_24} and \(k_M = 0.7\) \citep{d_Amico_2020}.  

\begin{table}
    \centering
    \begin{tabular}{c|c|c|c}
    \hline
        n & $b_i$ & $b_j$ & $P^n_l$\\ \hline
        1 & 1.0 & 1.0 & $P11l_1$\\ \hline
        2 & $b_1$ & 1.0 & $P11l_2$\\ \hline
        3 & $b_1$ & $b_1$ & $P11l_3$\\ \hline
        4 & 1.0 & 1.0 & $P\mathrm{loop}l_1$\\\hline
        5 & $b_1$ & 1.0 & $P\mathrm{loop}_2$\\ \hline
        6 & $b_2$ & 1.0 & $P\mathrm{loop}_3$\\ \hline
        7 & $b_3$ & 1.0 & $P\mathrm{loop}_4$\\ \hline
        8 & $b_4$ & 1.0 & $P\mathrm{loop}l_5$\\ \hline
        9 & $b_1$ & $b_1$ & $P\mathrm{loop}l_6$\\ \hline
        10 & $b_1$ & $b_2$ & $P\mathrm{loop}l_7$\\ \hline
        11 & $b_1$ & $b_3$ & $P\mathrm{loop}l_8$\\ \hline
        12 & $b_1$ & $b_4$ & $P\mathrm{loop}l_9$\\ \hline
        13 & $b_2$ & $b_2$ & $P\mathrm{loop}l_{10}$\\ \hline
        14 & $b_2$ & $b_4$ & $P\mathrm{loop}l_{11}$\\ \hline
        15 & $b_4$ & $b_4$ & $P\mathrm{loop}l_{12}$\\ \hline
        16 & $b_1$ & $c_{ct}$ & $2P_{\mathrm{lin}} \frac{k^2}{k_M^2}$\\ \hline
        17 & $b_1$ & $c_{r,1}$ & $2P_{\mathrm{lin}} \frac{k^2}{k_M^2}$\\ \hline
        18 & $b_1$ & $c_{r,2}$ & $2P_{\mathrm{lin}} \frac{k^2}{k_M^2}$\\ \hline
        19 & $c_{ct}$ & 1.0 & $2P_{\mathrm{lin}} \frac{k^2}{k_M^2}$\\ \hline
        20 & $c_{r,1}$ & 1.0 & $2P_{\mathrm{lin}} \frac{k^2}{k_M^2}$\\ \hline
        21 & $c_{r,2}$ & 1.0 & $2P_{\mathrm{lin}} \frac{k^2}{k_M^2}$\\ \hline
        22 & $c_{\epsilon, 1}$ & 1.0 & $\frac{1}{\overline{n_g}}$\\ \hline
        23 & $c_{\epsilon, \mathrm{mono}}$ & 1.0 & $\frac{1}{\overline{n_g}} \frac{k^2}{k_M^2}$\\ \hline
        24 & $c_{\epsilon, \mathrm{quad}}$ & 1.0 & $\frac{1}{\overline{n_g}} \frac{k^2}{k_M^2}$\\ \hline
    \end{tabular}
    \caption{The 24 different combinations of bias parameters in equation\ref{eq:LC}. The expressions for \(P11l\) and \(P\mathrm{loop}l\) are very long and not necessary to include in this paper. \(P_{\mathrm{lin}}\) here denotes the linear power spectrum from \textsc{CLASS}.}
    \label{tab:term_24}
\end{table}

\bsp	
\label{lastpage}
\end{document}